\documentclass[aps,prd,onecolumn,groupedaddress,showpacs,nofootinbib,amssymb]{revtex4}
\usepackage[dvips]{graphicx}
\usepackage{amssymb}
\usepackage{amsmath}
\usepackage{graphicx}
\usepackage{amsfonts}
\usepackage{bm}
\begin{document}

\title{Accelerating Cosmology and Phase Structure of F(R) Gravity with
Lagrange Multiplier Constraint: Mimetic Approach}

\author{S.D. Odintsov$^{1,2}$, V.K. Oikonomou$^{3,4,5}$}
\affiliation{$^1$Instituci\`{o} Catalana de Recerca i Estudis Avan\c{c}ats
(ICREA),
Barcelona, Spain \\
$^2$Institut de Ciencies de l'Espai (CSIC-IEEC), Campus UAB,
Campus UAB, Carrer de Can Magrans, s/n 08193 Cerdanyola del Valles,
Barcelona, Spain\\
$^{3)}$ Department of Theoretical Physics, Aristotle University of
Thessaloniki,
54124 Thessaloniki, Greece\\
$^{4)}$ Tomsk State Pedagogical University, 634061 Tomsk \\
$^{5)}$  Lab. Theor. Cosmology, Tomsk State University of Control Systems
and Radioelectronics (TUSUR), 634050 Tomsk, Russia
}

\begin{abstract}
We study mimetic $F(R)$ gravity with potential and Lagrange
multiplier constraint. In the context of these theories, we introduce a
reconstruction technique which enables us to realize arbitrary cosmologies,
given the Hubble rate and an arbitrarily chosen $F(R)$ gravity. We exemplify
our method by realizing cosmologies that are in concordance with current
observations (Planck data) and also well known bouncing cosmologies. The
attribute of our method is that the $F(R)$ gravity can be arbitrarily
chosen, so we can have the appealing features of the mimetic approach
combined with the known features of some $F(R)$ gravities, which unify
early-time with late-time acceleration. Moreover, we study the existence and
the stability of de Sitter points in the context of mimetic $F(R)$ gravity.
In the case of unstable de Sitter points, it is demonstrated that graceful
exit from inflation occurs. We also study the Einstein frame counterpart
theory of the Jordan frame mimetic $F(R)$ gravity, we discuss the general
properties of the theory and exemplify our analysis by studying a quite
interesting from a phenomenological point of view, model with two scalar
fields. We also calculate the observational indices of the two scalar field
model, by using the two scalar field formalism. Furthermore, we extensively
study the dynamical system that corresponds to the mimetic $F(R)$ gravity,
by finding the fixed points and studying their stability. Finally, we modify
our reconstruction method to function in the inverse way and thus yielding
which $F(R)$ gravity can realize a specific cosmological evolution, given
the mimetic potential and the Lagrange multiplier.
\end{abstract}

\pacs{95.35.+d, 98.80.-k, 98.80.Cq, 95.36.+x}

\maketitle

\section{Introduction}

One of the most sound cosmological observation in the last twenty years, was
the verification of a late-time acceleration of our Universe in the late
90's \cite{latetimeobse}. This observation has powered a vast research
stream aiming to model consistently this late-time acceleration, with
modified gravity playing a prominent role in this research stream. For
reviews on this vast subject consult \cite{reviews1} and for some insightful
studies on the subject see \cite{barrowearly}.
This nicely extends the universe evolution picture, taking into account that
early universe was also eventually accelerating (early-time
inflation\cite{inflrev}). The early-time inflation maybe also described in
terms of $F(R)$ gravity (for recent review, see \cite{bam1}). The most
appealing models of modified gravity pass  most of the tests posed by
observations and by local gravity constraints, and also offer the
possibility for a unified description of early-time and late-time
acceleration, within the same theoretical framework
\cite{sergeinojiri2003,reviews1}. The late-time acceleration is attributed
to a negative pressure perfect fluid called dark energy (for reviews see
\cite{reviews2}) which dominates the present Universe's energy density at a
comparably large percentage $\Omega_{DE}\sim 68.3\%$. The rest is controlled
by ordinary matter at a percentage $\Omega_m\sim4.9\%$ and by cold dark
matter with $\Omega_{DM}\sim
26.8\%$. With regards to the latter, as in the dark energy case, there is no
up to date experimentally or observationally confirmed theory which predicts
the nature of dark matter. It is speculated that it is of particle nature,
since in most extensions of the Standard Model \cite{darkmatter}, even in
the minimal supersymmetric ones \cite{shafi}, dark matter is treated as a
particle non-interacting with ordinary matter. Recently it was proposed by
Chamseddine and Mukhanov \cite{mukhanov1} that dark matter can be a result
of gravitational modification of General Relativity, the so-called mimetic
dark matter approach. The cosmological consequences were further studied in
\cite{mukhanov2,Golovnev} and extended in various theoretical frameworks in
\cite{NO2,mimetic1,mimetic2,mimetic3,mimeletter}. Actually Ref. \cite{mimeletter}, was a compact predecessor of the present work, so for a brief introduction to the issues we shall address here, consult \cite{mimeletter}. In their work \cite{mukhanov1},
Chamseddine and Mukhanov used a covariant approach in order to isolate the
conformal degree of freedom of the ordinary Einstein-Hilbert gravity. In the
final theory, the physical metric is defined in terms of an auxiliary scalar
field $\phi$, the first derivatives of which appear in the action. As a
result, the conformal degree of freedom is rendered dynamical, even in the
absence of ordinary matter fluids and it is exactly this conformal degree of
freedom that mimics cold dark matter. The theoretical framework of mimetic
gravity was extended as the mimetic $F(R)$ gravity case in Ref. \cite{NO2},
and the attribute of this mimetic $F(R)$ gravity approach is that it is
possible to harbor all the appealing properties of $F(R)$ gravity, such as
unification of early and late-time acceleration \cite{NO3}, and also have at
the same time a description of cold dark matter originating from the
gravitational sector, without the need of extra matter fluids to describe
it.

In this paper we shall study in detail the $F(R)$ gravity extension of
mimetic gravity approach, with scalar potential $V(\phi)$, to which we shall
refer as mimetic potential, and with Lagrange multiplier $\lambda (\phi)$.
By using the Lagrange multiplier formalism approach \cite{CMO} in Jordan
frame description,  we shall provide a reconstruction method which enables
us to investigate which mimetic potential and Lagrange multiplier can
realize an specific cosmological evolution, for an arbitrarily chosen $F(R)$
gravity. Since in the context of our reconstruction method the Hubble rate
and the $F(R)$ gravity can be arbitrarily chosen, it is possible to choose a
specific cosmological evolution, for which concordance with observations can
be achieved and also the $F(R)$ gravity can be chosen to be a viable $F(R)$
gravity in principle. We support our findings by using some illustrative
characteristic paradigms, for which we find the exact analytic forms of the
corresponding mimetic potential and Lagrange multiplier. In addition, we
thoroughly investigate the existence and stability of de Sitter points in
mimetic $F(R)$ gravity with potential and Lagrange multiplier, by checking
some phenomenologically interesting $F(R)$ models. We also study how certain
bouncing cosmologies and singular cosmologies can be realized in the context
of mimetic $F(R)$ gravity. Particularly, we investigate which $F(R)$
gravities can generate such a cosmological evolution, given the Hubble rate
of the cosmological evolution, the mimetic potential and the Lagrange
multiplier, which can be arbitrarily chosen. The resulting picture is quite
interesting since the same cosmological evolution can be realized from
different $F(R)$ gravities, which correspond to different mimetic potentials
and Lagrange multipliers. In addition, we study the Einstein frame
counterpart theory of the mimetic $F(R)$ gravity with potential and Lagrange
multiplier, exemplifying the general theoretical construction by using the
Einstein frame counterpart of the Jordan frame $R^2$ gravity. As we
demonstrate, the resulting theory is quite interesting since it is quite
similar to the two scalar model appearing in supersymmetric frameworks, as
in Ref. \cite{linde}. Also for the resulting two scalar model we calculate
in detail the corresponding observational indices, using very well known
techniques for the two scalar field indices \cite{kaizer,twoscalarindices}.
Finally, we perform a thorough analysis of the dynamical system
corresponding to mimetic $F(R)$ gravity, focusing on finding the fixed
points of the dynamical system and studying their stability. Moreover, we
also discuss the physical significance of the fixed points as unification of
inflation with dark energy and the possibility of connecting some of the
fixed points via some attractor that leads from one to another.

We need to stress that the study we performed should be considered by also taking into account the observational limits for the generic picture from cosmological/astronomical surveys. Particularly, in a future work our results regarding certain cosmological scenarios, should be enriched with a study which will take into account certain observational constraints, as for example was done in \cite{Cai:2014bea} in the case of non-singular bouncing cosmologies. Such theoretical study is compelling in order to render the present theoretical study more complete and sound from a theoretical point of view, and we intend to address such issues in a future work.

 This paper is organized as follows: In section II we present the
reconstruction method, which enable us to find which mimetic potential and
Lagrange multiplier generate an arbitrary cosmological evolution, given the
$F(R)$ gravity. In section III, we investigate the existence and stability
of de Sitter vacua, in the context of mimetic $F(R)$ gravity and we discuss
the significance of the existence of unstable de Sitter points in connection
with inflation, and the possible  graceful exit from inflation. In section
IV,
 we realize several bouncing cosmologies, in the context of mimetic $F(R)$
gravity. Particularly we investigate which $F(R)$ gravity can realize the
bouncing cosmologies under study, given the mimetic potential and the
Lagrange multiplier which are arbitrarily chosen. As we shall demonstrate,
the resulting $F(R)$ gravities are quite different from the ones we obtain
in the pure $F(R)$ gravity case. In section V we study the Einstein frame
counterpart theory of the Jordan frame mimetic $F(R)$ gravity. We specify
our analysis by choosing a well known $F(R)$ gravity model in the Jordan
frame and we discuss the implications of this model in the Einstein frame.
In addition, we calculate the observational indices of the resulting two
scalar field model, by using the quite general multi-scalar field
approximation. In section VI, we thoroughly study the dynamical system that
the mimetic $F(R)$ gravity corresponds to. After introducing some
conveniently chosen variables, we investigate which are the fixed points of
the dynamical system and also study their stability against linear
perturbations. We also investigate if there exist attractor solutions that
can connect dynamically some of the fixed points. Finally, in section VII,
we study which $F(R)$ gravities can generate the cosmological evolutions
presented in section II, given the mimetic potential and the Lagrange
multiplier. The conclusions are presented at the end of the paper.

\section{Concordance with Observations using Lagrange-multiplier Mimetic
$F(R)$ Gravity}

In mimetic $F(R)$ gravity the conformal symmetry is a non-violated internal
degree of freedom \cite{mukhanov1}. Following \cite{mukhanov1}, the physical
metric $g_{\mu \nu}$ is written in terms of an auxiliary scalar field $\phi$
and an auxiliary metric tensor $\hat{g}_{\mu \nu}$, as follows,
\begin{equation}\label{metrpar}
g_{\mu \nu}=-\hat{g}^{\mu \nu}\partial_{\rho}\phi \partial_{\sigma}\phi
\hat{g}_{\mu \nu}\, .
\end{equation}
Instead of varying the gravitational action with respect to the metric
$g_{\mu \nu}$, in Ref. \cite{mukhanov1} the authors considered variation
with respect to the metric $\hat{g}_{\mu \nu}$ and with respect to the
auxiliary scalar field $\phi$. It follows from Eq. (\ref{metrpar}) that,
\begin{equation}\label{impl1}
g^{\mu \nu}(\hat{g}_{\mu \nu},\phi)\partial_{\mu}\phi\partial_{\nu}\phi=-1\,
.
\end{equation}
The Weyl transformation $\hat{g}_{\mu \nu}=e^{\sigma (x)}g_{\mu \nu}$ leaves
the parametrization (\ref{metrpar}) invariant, and the metric $\hat{g}_{\mu
\nu}$ does not appear in the final action. We shall assume that the physical
metric $g_{\mu \nu}$ is the following,
\begin{equation}\label{frw}
ds^2 = - dt^2 + a(t)^2 \sum_{i=1,2,3}
\left(dx^i\right)^2\, ,
\end{equation}
which is flat Friedmann-Robertson-Walker (FRW), with the parameter $a(t)$
denoting the scale factor, and we also assume that the auxiliary scalar
field $\phi$ depends only on the cosmic time $t$. For a flat FRW metric, the
Ricci scalar is equal to $R=6\left (\dot{H}+2H^2 \right )$, with $H(t)$
denoting the Hubble rate. The mimetic $F(R)$ theory with a scalar potential
$V(\phi )$ and a Lagrange multiplier $\lambda (\phi )$ in the Jordan frame
has the following action\cite{NO2},
\begin{equation}\label{actionmimeticfraction}
S=\int \mathrm{d}x^4\sqrt{-g}\left ( F\left(R(g_{\mu \nu})\right
)-V(\phi)+\lambda \left(g^{\mu \nu}\partial_{\mu}\phi\partial_{\nu}\phi
+1\right)\right )\, .
\end{equation}
As it can be seen, we assumed that no matter fluids are present in the
action and this is what we assume for the rest of this paper. Upon variation
of the action (\ref{actionmimeticfraction}) with respect to the physical
metric $g_{\mu \nu}$, we get,
\begin{align}\label{aeden}
& \frac{1}{2}g_{\mu \nu}F(R)-R_{\mu
\nu}F'(R)+\nabla_{\mu}\nabla_{\nu}F'(R)-g_{\mu \nu}\square F'(R)\\ \notag &
\frac{1}{2}g_{\mu \nu}\left (-V(\phi)+\lambda \left( g^{\rho
\sigma}\partial_{\rho}\phi\partial_{\sigma}\phi+1\right) \right )-\lambda
\partial_{\mu}\phi \partial_{\nu}\phi =0 \, .
\end{align}
In addition, upon variation of the action (\ref{actionmimeticfraction}),
with respect to the scalar field $\phi$, we get,
\begin{equation}\label{scalvar}
-2\nabla^{\mu} (\lambda \partial_{\mu}\phi)-V'(\phi)=0\, ,
\end{equation}
with the ``prime'' here denoting differentiation with respect to the scalar
field $\phi$. Varying the action (\ref{actionmimeticfraction}) with respect
to the Lagrange multiplier $\lambda$, we obtain,
\begin{equation}\label{lambdavar}
g^{\rho \sigma}\partial_{\rho}\phi\partial_{\sigma}\phi=-1\, ,
\end{equation}
a result which is identical to the one of Eq. (\ref{impl1}). For a flat FRW
background, since the auxiliary scalar field depends only on the cosmic time
$t$, we may rewrite the equations of motion (\ref{aeden}), (\ref{scalvar})
and (\ref{lambdavar}), in the following way,
\begin{equation}\label{enm1}
-F(R)+6(\dot{H}+H^2)F'(R)-6H\frac{\mathrm{d}F'(R)}{\mathrm{d}t}-\lambda
(\dot{\phi}^2+1)+V(\phi)=0\, ,
\end{equation}
\begin{equation}\label{enm2}
F(R)-2(\dot{H}+3H^2)+2\frac{\mathrm{d}^2F'(R)}{\mathrm{d}t^2}+4H\frac{\mathrm{d}F'(R)}{\mathrm{d}t}-\lambda (\dot{\phi}^2-1)-V(\phi)=0\, ,
\end{equation}
\begin{equation}\label{enm3}
2\frac{\mathrm{d}(\lambda \dot{\phi})}{\mathrm{d}t}+6H\lambda
\dot{\phi}-V'(\phi)=0\, ,
\end{equation}
\begin{equation}\label{enm4}
\dot{\phi}^2-1=0\, ,
\end{equation}
with the ``dot'' denoting differentiation with respect to the time $t$,
while this time, the ``prime'' in Eqs. (\ref{enm1}) and (\ref{enm2})
indicates differentiation with respect to the scalar curvature $R$. Also in
Eq. (\ref{enm3}) the prime denotes differentiation with respect to the
auxiliary scalar field $\phi$.

It easily follows from Eq. (\ref{enm3}) that the auxiliary scalar field
$\phi$ is identified with the cosmic time $t$. Note that the same
identification was possible in the Einstein-Hilbert mimetic gravity
theoretical framework, as a consequence of Eq. (\ref{impl1}), consult Ref.
\cite{mukhanov1} for more details on this. In view of the identification
$t=\phi$, Eq. (\ref{enm2}) can be rewritten as follows,
\begin{equation}\label{sone}
F(R)-2(\dot{H}+3H^2)+2\frac{\mathrm{d}^2F'(R)}{\mathrm{d}t^2}+4H\frac{\mathrm{d}F'(R)}{\mathrm{d}t}-V(t)=0\, ,
\end{equation}
and therefore the mimetic potential $V(\phi=t)$ that generates the
cosmological evolution with Hubble rate $H(t)$, for a given $F(R)$ gravity,
is equal to,
\begin{equation}\label{scalarpot}
V(\phi=t)=2\frac{\mathrm{d}^2F'(R)}{\mathrm{d}t^2}+4H\frac{\mathrm{d}F'(R)}{
\mathrm{d}t}+F(R)-2(\dot{H}+3H^2)\, .
\end{equation}
Combining Eqs. (\ref{scalarpot}) and (\ref{enm1}), we get the Lagrange
multiplier $\lambda (t)$, which is,
\begin{equation}\label{lagrange}
\lambda (t)=-3 H \frac{\mathrm{d}F'(R)}{\mathrm{d}t}+3
(\dot{H}+H^2)-\frac{1}{2}F(R)\, .
\end{equation}
Practically, we have at hand a reconstruction method which enables us to
find the scalar potential and the Lagrange multiplier that can generate a
cosmological evolution with Hubble rate $H(t)$, given the, in principle,
arbitrary $F(R)$ gravity. A very useful differential equation can be
obtained if we combine Eqs. (\ref{scalarpot}) and (\ref{lagrange}), and it
is the following,
\begin{equation}\label{diffeqnnewaddition}
2\frac{\mathrm{d}^2F'(R)}{\mathrm{d}t^2}+4H\frac{\mathrm{d}F'(R)}{\mathrm{d}
t}- 6 H \frac{\mathrm{d}F'(R)}{\mathrm{d}t}-2 (\dot{H}+3
H^2)+6(\dot{H}+H^2)\frac{\mathrm{d}F'(R)}{\mathrm{d}t}-2 \lambda
(t)-V(t)=0\, \end{equation}
which we shall frequently use in the sections to follow. The novelty of the
reconstruction method we propose here is that by choosing any, viable or
not, $F(R)$ gravity, we can produce any cosmological evolution we wish,
which can be chosen in such a way so that it is concordance with
observational data. The freedom in this reconstruction method is provided by
the mimetic potential and the Lagrange multiplier, at the cost that these
might have a complicated form. In the rest of this section we exemplify our
reconstruction method by using some characteristic examples. For a recent account on exactly this issue, see \cite{mimeletter}.

We shall consider a general class of $F(R)$ gravity models with the $F(R)$
gravity function being of the following form,
\begin{equation}\label{generalfrmodels}
F(R)=R+\mu R^p+d R^q\, ,
\end{equation}
with $d,f,p,q$ being arbitrary parameters. This class of models is known to
provide quite rich phenomenology, with regards to late and early time
acceleration, see \cite{reviews1} for more details. A model belonging to the
class of models (\ref{generalfrmodels}), with profound astrophysical
significance is the following,
\begin{equation}\label{staro}
F(R)=R-d R^3+f R^2\, ,
\end{equation}
with the parameters $d$ and $f$ being positive real numbers, arbitrarily
chosen. This model has quite appealing phenomenological features, since as
was shown in Ref. \cite{sergeicappostars}, when this model is employed to
neutron stars, it generates an increase in the maximal neutron star mass,
which is a very interesting phenomenological feature. Our purpose is to find
which mimetic scalar potential $V(\phi=t)$ and Lagrange multiplier generates
a cosmological evolution compatible with observations, for the $F(R)$
gravity of Eq. (\ref{staro}). We shall employ the formalism for the
calculation of the Jordan frame observational indices, which was developed
in Ref. \cite{sergeislowroll} (see also \cite{muk}). As was demonstrated in
\cite{sergeislowroll}, the $F(R)$ gravity can be treated as a perfect fluid,
and in effect the observational indices can be expressed in terms of the
Hubble rate and its higher derivatives. The same applies in our case so we
refrain from going into details, however we need to stress that this
formalism is valid if some slow-roll approximation is used, meaning that the
slow-roll indices have small values during inflation. Consequently, our aim
is to investigate which mimetic potential and Lagrange multiplier can
generate a viable cosmological evolution. We shall express the observational
indices and all physical quantities we shall use as functions of the
$e$-folding number $N$, instead of the cosmic time $t$. Bearing in mind that
the following transformations,
\begin{equation}\label{transfefold}
\frac{\mathrm{d}}{\mathrm{d}t}=H(N)\frac{\mathrm{d}}{\mathrm{d}N},{\,}{\,}{\
,}
\frac{\mathrm{d}^2}{\mathrm{d}t^2}=H^2(N)\frac{\mathrm{d}^2}{\mathrm{d}N^2}+
H(N)\frac{\mathrm{d}H}{\mathrm{d}N}\frac{\mathrm{d}}{\mathrm{d}N}\, ,
\end{equation}
and according to Ref. \cite{sergeislowroll}, the slow-roll indices are
written as functions of the $e$-folding number $N$, in the following way,
\begin{align}\label{hubbleslowrollnfolding}
&\epsilon=-\frac{H(N)}{4 H'(N)}\left(\frac{\frac{H''(N) }{H(N)}+6\frac{H'(N)
}{H(N)}+\left(\frac{H'(N)}{H(N)}\right)^2}{3+\frac{H'(N)}{H(N)}}\right)^2
\, ,\\ \notag &
\eta=-\frac{\left(9\frac{H'(N)}{H(N)}+3\frac{H''(N)}{H(N)}+\frac{1}{2}\left(
\frac{H'(N)}{H(N)}\right)^2-\frac{1}{2}\left(
\frac{H''(N)}{H'(N)}\right)^2+3
\frac{H''(N)}{H'(N)}+\frac{H'''(N)}{H'(N)}\right)}{2\left(3+\frac{H'(N)}{H(N
)}\right)}\, ,
\end{align}
with the prime this time denoting differentiation with respect to $N$.
According to the formalism of Ref. \cite{sergeislowroll}, the  spectral
index of primordial curvature perturbations and the scalar-to-tensor ratio
$r$ read,
\begin{equation}\label{indexspectrscratio}
n_s\simeq 1-6 \epsilon +2\eta,\, \, \, r=16\epsilon \, .
\end{equation}
As we already mentioned, these relations are valid, as long as the slow-roll
indices satisfy $\epsilon,\eta \ll 1$. Having the analytical expressions for
the slow-roll indices at hand, we can choose a compatible to the
observational data evolution and find the potential $V(N)$ and the Lagrange
multiplier $\lambda (N)$, that generate this cosmological evolution, given
the $F(R)$ gravity of Eq. (\ref{staro}). According to the recent Planck
observational data \cite{planck}, the aforementioned observational indices
are constrained as follows,
\begin{equation}\label{constraintedvalues}
n_s=0.9644\pm 0.0049\, , \quad r<0.10\, .
\end{equation}
Let us for example the cosmological evolution, with the following Hubble
rate,
\begin{equation}\label{hub2}
H(N)=\left(-G_0\text{  }e^{\beta N }+G_1\right)^b\, ,
\end{equation}
and by using Eqs. (\ref{hubbleslowrollnfolding}), the parameter $\epsilon$
is equal to,
\begin{align}\label{hubslowroll2}
& \epsilon=-\frac{b e^{\beta N} G_0 \beta  \left(G_1 (6+\beta )-2 e^{\beta
N} G_0 (3+b \beta )\right)^2}{4 \mathcal{F}(N)}
\end{align}
where the function $\mathcal{F}(N)$ is equal to,
\begin{equation}\label{hfghfhgfghdf}
\mathcal{F}(N)=\left(e^{\beta N} G_0-G_1\right) \left(-3 G_1+e^{\beta N} G_0
(3+b \beta )\right)^2
\end{equation}
while $\eta$ is equal to,
\begin{equation}\label{edggs}
\eta =-\frac{\beta  \left(8 b^2 e^{2\beta N} G_0^2 \beta +G_1 \left(2
e^{\beta N} G_0 (-3+\beta )+G_1 (6+\beta )\right)+2 b e^{\beta N} G_0
\left(12 e^{\beta N} G_0-G_1 (12+5 \beta )\right)\right)}{4 \left(e^{\beta
N} G_0-G_1\right) \left(-3 G_1+e^{\beta N} G_0 (3+b \beta )\right)}
\, .
\end{equation}
Using Eqs. (\ref{hubslowroll2}) and (\ref{edggs}), the spectral index and
scalar-to-tensor ratio can be calculated, and $n_s$ reads,
\begin{align}\label{scalarpertandsctotenso}
& n_s=\frac{2 \left(e^N\right)^{3 \beta } G_0^3 (3+b \beta )^2 (1+2 b \beta
)+3 G_1^3 \left(-6+6 \beta +\beta ^2\right)}{2 \mathcal{F}(N)}
+\frac{e^{\beta N} G_0 G_1^2 \left(54+12 (-3+4 b) \beta +3 \beta ^2+2 b
\beta ^3\right)}{2 \mathcal{F}(N)}\\ \notag &
-\frac{2 e^{2\beta N} G_0^2 G_1 \left(27+(-9+48 b) \beta +\left(3+13
b^2\right) \beta ^2+b (1+b) \beta ^3\right)}{2 \mathcal{F}(N)}
\, ,
\end{align}
while the scalar-to-tensor is equal to,
\begin{equation}\label{thodorakis}
r=-\frac{4 b e^{\beta N} G_0 \beta  \left(G_1 (6+\beta )-2 e^{\beta N} G_0
(3+b \beta )\right)^2}{\mathcal{F}(N)}
\end{equation}
In principle, there are many values of the parameters $G_0$, $G_1$, $\beta$
and $b$, for which concordance with the observations can be achieved, like
for example $G_0=0.5$, $G_1=12$, $\beta=0.024$ and $b=1/2$, but a convenient
choice is the following,
\begin{equation}\label{parmchoice12}
G_0=0.5,\,\,\, G_1=10,\,\,\,\beta=0.0222,\,\,\, b=1\, ,
\end{equation}
for which, the indices $n_s$ and $r$ become equal to,
\begin{equation}\label{indnewparadigm12}
n_s\simeq 0.96567, \,\,\, r=0.0640848\, ,
\end{equation}
which are compatible with the Planck data (\ref{constraintedvalues}). The
reason for choosing the values for the parameters as in Eq.
(\ref{parmchoice12}), is that for these it is easy to express the resulting
potential $V(N)$ as a function of $\phi$. Using the function $G(N)=H(N)^2$
(see also \cite{sergeirecon}), the Ricci scalar as a function of $G(N)$
reads,
\begin{equation}\label{cnc}
R(N)=3G'(N)+12G(N)\, .
\end{equation}
Combining Eqs. (\ref{staro}), (\ref{hub2}), (\ref{cnc}), and
(\ref{scalarpot}), we obtain the mimetic potential $V(N)$ for the $F(R)$
gravity of Eq. (\ref{staro}), which reads,
\begin{align}\label{podnegsa}
& V(N)=
\mathcal{S}(N)^{2 b} \Big{(}6+3 \mathcal{S}(N)^{2 b}+9 f \mathcal{S}(N)^{2
b} \Big{(}4+\mathcal{S}(N)^{2 b}\Big{)}^2-27 d \mathcal{S}(N)^{4 b}
\Big{(}4+\mathcal{S}(N)^{2 b}\Big{)}^3
\\ \notag & +8 \Big{(}f-9 d \mathcal{S}(N)^{2 b} \Big{(}4+\mathcal{S}(N)^{2
b}\Big{)}\Big{)}+\frac{2 b e^{\beta N} G_0 \beta }{\mathcal{S}(N)}-48 b
e^{\beta N} G_0 \mathcal{S}(N)^{2 (-1+b)} \beta  \Big{(}f \Big{(}G_1
\Big{(}2+\mathcal{S}(N)^{2 b}\Big{)} (1+\beta ) \\ \notag & -e^{\beta N} G_0
\Big{(}2+\mathcal{S}(N)^{2 b}+4 b \Big{(}1+\mathcal{S}(N)^{2 b}\Big{)} \beta
\Big{)}\Big{)}+9 d \mathcal{S}(N)^{2 b} \Big{(}-G_1 \Big{(}8+6
\mathcal{S}(N)^{2 b}+\mathcal{S}(N)^{4 b}\Big{)} (1+\beta )
\\ \notag & +e^{\beta N} G_0 \Big{(}8+6 \mathcal{S}(N)^{2
b}+\mathcal{S}(N)^{4 b}+4 b \Big{(}8+9 \mathcal{S}(N)^{2 b}+2
\mathcal{S}(N)^{4 b}\Big{)} \beta \Big{)}\Big{)}\Big{)}\Big{)} \, ,
\end{align}
where $\mathcal{S}(N)=-e^{N\beta } G_0+G_1$. In the same way, by combining
Eqs. (\ref{staro}), (\ref{hub2}), (\ref{cnc}) and (\ref{lagrange}), we
obtain the Lagrange multiplier, which is,
\begin{align}\label{lajdbfe}
& \lambda (N)=
\frac{3}{2} \mathcal{S}(N)^{2 b} \Big{(}-4-\mathcal{S}(N)^{2 b}-3 f
\mathcal{S}(N)^{2 b} \Big{(}4+\mathcal{S}(N)^{2 b}\Big{)}^2+9 d
\mathcal{S}(N)^{4 b} \Big{(}4+\mathcal{S}(N)^{2 b}\Big{)}^3\\ \notag & +4
\Big{(}-f+9 d \mathcal{S}(N)^{2 b} \Big{(}4+\mathcal{S}(N)^{2
b}\Big{)}\Big{)}\\ \notag & +\frac{2 \Big{(}-1-6 f \mathcal{S}(N)^{2 b}
\Big{(}4+\mathcal{S}(N)^{2 b}\Big{)}+27 d \mathcal{S}(N)^{4 b}
\Big{(}4+\mathcal{S}(N)^{2 b}\Big{)}^2\Big{)} \Big{(}-G_1+e^{\beta N} G_0
(1+b \beta )\Big{)}}{\mathcal{S}(N)}\Big{)}
 \, .
\end{align}

Using the values of the parameters as in Eq. (\ref{parmchoice12}),we can
express the cosmic time $t$ as a function of the scale factor $\alpha$, as
follows: since $H(N(\alpha))$ is equal to,
\begin{equation}\label{hna}
H(N(\alpha))=\left(-G_0\text{  }(\alpha /a_0)^{\beta }+G_1\right)^b\, ,
\end{equation}
and since that $e^N=\frac{\alpha}{\alpha_0}$, integrating the above equation
and using the values of the parameters given in Eq. (\ref{parmchoice12})
($b=1$ is the most relevant for this calculation), we get the function
$t=t(\alpha)$, which reads,
\begin{equation}\label{ta}
t+c_1=\frac{\beta  \ln(\alpha )-\ln\left(G_1-G_0 \left(\frac{\alpha
}{a_0}\right)^{\beta }\right)}{G_1 \beta }\, .
\end{equation}
This equation can explicitly be solved with respect to the scale factor
$\alpha$,
\begin{equation}\label{scalef}
\alpha (t)= \left
(\frac{e^{G_1\beta(t+c_1)}G_1}{1+e^{G_1\beta(t+c_1)}\frac{G_0}{\alpha_0^{\beta}}}\right)^{\frac{1}{\beta}},\,\,\,c_1=\frac{1}{\beta G_1}\ln
\left(\frac{\alpha_0}{G_1-\frac{G_0}{\alpha_0^{\beta-1}}}\right)\, .
\end{equation}
Bearing in mind that $\phi=t$, by substituting the following in Eq.
(\ref{podnegsa}),
\begin{equation}\label{finalsubstitution}
e^{\beta N} \rightarrow
\frac{e^{G_1\beta(\phi+c_1)}G_1}{1+e^{G_1\beta(\phi+c_1)}\frac{G_0}{\alpha_0
^{\beta}}}\, ,
\end{equation}
we get the scalar potential $V(\phi)$. Hence, as we demonstrated, by using
the reconstruction method we propose, it is possible to realize a large
number of cosmologies which are compatible with observations, by making use
of an arbitrarily chosen $F(R)$ gravity. This arbitrariness of the choice of
the $F(R)$ gravity is the novel feature of our approach. For illustrative
reasons we used an appealing model of $F(R)$ gravity with phenomenological
significance, that of Eq. (\ref{staro}), but in principle other models may
be used, for example viable models of $F(R)$ gravity, like the ones used in
\cite{reviews1}. If someone uses viable models of modified gravity in the
context of the mimetic $F(R)$ gravity theoretical framework we propose, then
it is possible to have all the attributes that our formalism brings along,
combined with all the appealing features of viable modified gravity models,
for example the unification of early time with late-time acceleration.

For illustrative purposes we shall study another model of modified gravity,
which is a viable model, the Hu-Sawicki model \cite{husawicki}, the $F(R)$
function of which is,
\begin{equation}\label{husawi}
F(R)=R-\mu  R_c \frac{(R/R_c)^{2n}}{(R/R_c)^{2n}+1}
\end{equation}
where $R_c$ and $\mu $ are positive real phenomenological parameters
constrained by observations \cite{husawicki}, while $n$ is a positive real
number. Also let us consider the class of cosmological models with the
Hubble rate being of the following form,
\begin{equation}\label{hub1}
H(N)=\left(-G_0\text{  }N^{\beta }+G_1\right)^b\, ,
\end{equation}
where $G_0$, $G_1$, $\beta$ and $b$ are real numbers arbitrary in general.
In order to avoid negative values of the Hubble rate, the parameter $b$ must
be chosen in such a way so that it is of the form $b=\frac{2n}{2m+1}$, with
$m$ and $n$ positive integers, and also $b$ must obey the inequality $b<1$.
Substituting the Hubble rate (\ref{hub1}), to the slow-roll parameters of
Eq. (\ref{hubbleslowrollnfolding}), the parameter $\epsilon$ reads,
\begin{align}\label{hubpar1}
& \epsilon=N^{1-\beta } \Big{(}G_1-G_0 N^{\beta }\Big{)}\Big{(}3-\frac{b G_0
N^{-1+\beta } \beta }{G_1-G_0 N^{\beta }}\Big{)}^{-2}(4 b G_0 \beta )^{-1}
\Big{(}-\frac{6 b G_0 N^{-1+\beta } \beta }{G_1-G_0 N^{\beta }}+\frac{b^2
G_0^2 N^{-2+2 \beta } \beta ^2}{\Big{(}G_1-G_0 N^{\beta }\Big{)}^2} \\
\notag &
+\Big{(}G_1-G_0 N^{\beta }\Big{)}^{-b} \Big{(}-b G_0 N^{-2+\beta }
\Big{(}G_1-G_0 N^{\beta }\Big{)}^{-1+b} (-1+\beta ) \beta  \\ \notag &
+(-1+b) b G_0^2 N^{-2+2 \beta } \Big{(}G_1-G_0 N^{\beta }\Big{)}^{-2+b}
\beta ^2\Big{)}\Big{)}^2 \, .
\end{align}
and correspondingly the parameter $\eta$ reads,
\begin{align}\label{etaparameterversion1}
& \eta=-\frac{G_1^2 (-1+\beta ) (-3+6 N+\beta )+G_0^2 N^{2 \beta }
\left(3-10 b \beta +8 b^2 \beta ^2+6 N (-1+4 b \beta )\right)}{4 N
\left(G_1-G_0 N^{\beta }\right) \left(3 G_1 N-G_0 N^{\beta } (3 N+b \beta
)\right)}\\ \notag &
+\frac{-2 G_0 G_1 N^{\beta } (3 N (-2+\beta +4 b \beta )+(-1+\beta )
(-3+(-1+5 b) \beta ))}{4 N \left(G_1-G_0 N^{\beta }\right) \left(3 G_1 N-G_0
N^{\beta } (3 N+b \beta )\right)}
\end{align}
Then by using Eqs. (\ref{hubpar1}) and (\ref{etaparameterversion1}), the
observational parameters of Eq. (\ref{indexspectrscratio}) read,
\begin{align}\label{observa}
& n_s=\frac{1}{2 N \Big{(}G_1-G_0 N^{\beta }\Big{)} \Big{(}-3 G_1 N+G_0
N^{\beta } (3 N+b \beta )\Big{)}^2}\times \\ \notag &
\Big{(}3 G_1^3 N \Big{(}-3+6 N (1+N)+4 \beta -6 N \beta -\beta
^2\Big{)}-G_0^3 N^{3 \beta } (9 N (-1+2 N (1+N)) \\ \notag &
 +48 b N^2 \beta +2 b^2 (-1+13 N) \beta ^2+4 b^3 \beta ^3\Big{)}-G_0 G_1^2
N^{\beta } \Big{(}54 N^3+2 b (-1+\beta ) \beta ^2+3 N (-1+\beta ) (9+\beta )
\\ \notag &  +6 N^2 (9-6 \beta +8 b \beta )\Big{)}+G_0^2 G_1 N^{2 \beta }
\Big{(}54 N^3+2 b (1+b) (-1+\beta ) \beta ^2 \\ \notag &
 +6 N^2 (9-3 \beta +16 b \beta )+N \Big{(}-27+2 \beta  \Big{(}6+3 \beta +13
b^2 \beta \Big{)}\Big{)}\Big{)}\Big{)} \, .
\end{align}
By making the following choice for the free parameters $G_0$, $G_1$, $\beta$
and $b$,
\begin{equation}\label{defparam}
G_0=0.00005,\, \, \, G_1=3000, \, \, \,\beta=3, \, \, \, b=\frac{8}{9}\, ,
\end{equation}
we can calculate the spectral index of primordial curvature perturbations
and the scalar-to-tensor ratio, which for $N=60$ $e$-foldings
become equal to,
\begin{equation}\label{scalindex}
n_s\simeq 0.966157, \, \, \, r\simeq 0.000258947\, ,
\end{equation}
which are in concordance with the constraints (\ref{constraintedvalues}) of
the Planck mission \cite{planck}. Having established that the cosmological
evolution of Eq. (\ref{hub1}), is compatible with the latest observational
data, we may easily find which mimetic potential and Lagrange multiplier can
realize such a cosmological evolution, for the $F(R)$ gravity being that of
Eq. (\ref{husawi}). Combining Eqs. (\ref{hub1}), (\ref{husawi}),
(\ref{cnc}), and (\ref{scalarpot}), we get the mimetic scalar potential
$V(N)$, which reads,
\begin{align}\label{explicitpot1}
& V(N)=12 \mathcal{K}(N)^{2 b}+3 \mathcal{K}(N)^{4
b}+\frac{\mathcal{K}(N)^{-1+2 b} \Big{(}-6 G_1 N+2 G_0 N^{\beta } (3 N+b
\beta )\Big{)}}{N}\\ \notag & -\frac{9^n \mathcal{F}(N)^{2 n} R_c \mu
}{1+9^n \mathcal{F}(N)^{2 n}}+\frac{32 b G_0 n N^{-1+\beta }
\Big{(}2+\mathcal{K}(N)^{2 b}\Big{)} \Big{(}1-2 n+\mathcal{V}(N)^{2 n}+2 n
\mathcal{V}(N)^{2 n}\Big{)} \mathcal{V}(N)^{2 n} R_c \beta  \mu }{3
\Big{(}-G_1+G_0 N^{\beta }\Big{)} \Big{(}4+\mathcal{K}(N)^{2 b}\Big{)}^2
\Big{(}1+\mathcal{V}(N)^{2 n}\Big{)}^3}
\\ \notag & +\Big{(}16\ 3^{-1+2 n} b G_0 n N^{-2+\beta } \mathcal{F}(N)^{2
n} R_c \beta  \Big{(}-N \mathcal{K}(N) \Big{(}2+\mathcal{K}(N)^{2 b}\Big{)}
\Big{(}4+\mathcal{K}(N)^{2 b}\Big{)} \Big{(}1+9^n \mathcal{F}(N)^{2
n}\Big{)} \times  \\ \notag & \Big{(}1+2 n \Big{(}-1+9^n \mathcal{F}(N)^{2
n}\Big{)}+9^n \mathcal{F}(N)^{2 n}\Big{)}-G_1 \Big{(}8+6 \mathcal{K}(N)^{2
b}+\mathcal{K}(N)^{4 b}\Big{)} \Big{(}1+9^n \mathcal{F}(N)^{2
n}\Big{)}\times
\\ \notag &  \Big{(}1+2 n \Big{(}-1+9^n \mathcal{F}(N)^{2 n}\Big{)}+9^n
\mathcal{F}(N)^{2 n}\Big{)} (-1+\beta )-G_0 N^{\beta } \Big{(}16 b n^2
\Big{(}2+\mathcal{K}(N)^{2 b}\Big{)}^2 \Big{(}1-4\ 9^n \mathcal{F}(N)^{2
n}+81^n \mathcal{F}(N)^{4 n}\Big{)} \beta \\ \notag & +\Big{(}1+9^n
\mathcal{F}(N)^{2 n}\Big{)}^2 \Big{(}8+6 \mathcal{K}(N)^{2
b}+\mathcal{K}(N)^{4 b}+4 b \Big{(}4+3 \mathcal{K}(N)^{2
b}+\mathcal{K}(N)^{4 b}\Big{)} \beta \Big{)}\\ \notag & +2 n \Big{(}-1+81^n
\mathcal{F}(N)^{4 n}\Big{)} \Big{(}8+6 \mathcal{K}(N)^{2
b}+\mathcal{K}(N)^{4 b}+4 b \Big{(}8+7 \mathcal{K}(N)^{2 b}+2
\mathcal{K}(N)^{4 b}\Big{)} \beta \Big{)}\Big{)}\Big{)} \mu \Big{)}\times \\
\notag & \Big{(}\mathcal{K}(N)^2 \Big{(}4+\mathcal{K}(N)^{2 b}\Big{)}^3
\Big{(}1+9^n \mathcal{F}(N)^{2 n}\Big{)}^4\Big{)}^{-1}
\, ,
\end{align}
with $\mathcal{K}(N)=G_1-G_0 N^{\beta }$ and also with the function
$\mathcal{F}(N)$ being equal to,
\begin{equation}\label{ffynction}
\mathcal{F}(N)=\frac{\left(G_1-G_0 N^{\beta }\right)^{2 b}
\left(4+\left(G_1-G_0 N^{\beta }\right)^{2 b}\right)}{R_c}\, .
\end{equation}
In addition, the corresponding Lagrange multiplier reads,
\begin{align}\label{sder}
& \lambda (N)=
\frac{8\ 9^n b G_0 n N^{-1+\beta } \Big{(}2+\mathcal{K}(N)^{2 b}\Big{)}
\Big{(}1+2 n \Big{(}-1+9^n \mathcal{F}(N)^{2 n}\Big{)}+9^n \mathcal{F}(N)^{2
n}\Big{)} \mathcal{F}(N)^{2 n} R_c \beta  \mu }{\mathcal{K}(N)
\Big{(}4+\mathcal{K}(N)^{2 b}\Big{)}^2 \Big{(}1+9^n \mathcal{F}(N)^{2
n}\Big{)}^3}\\ \notag & +\frac{\Big{(}G_1 N-G_0 N^{\beta } (N+b \beta
)\Big{)} \Big{(}3 \mathcal{K}(N)^{2 b} \Big{(}4+\mathcal{K}(N)^{2 b}\Big{)}
\Big{(}1+9^n \mathcal{F}(N)^{2 n}\Big{)}^2-2\ 9^n n \mathcal{F}(N)^{2 n} R_c
\mu \Big{)}}{N \mathcal{K}(N) \Big{(}4+\mathcal{K}(N)^{2 b}\Big{)}
\Big{(}1+9^n \mathcal{F}(N)^{2 n}\Big{)}^2}\\ \notag & +\frac{1}{2}
\Big{(}-3 \mathcal{K}(N)^{2 b} \Big{(}4+\mathcal{K}(N)^{2
b}\Big{)}+\frac{9^n \mathcal{F}(N)^{2 n} R_c \mu }{1+9^n \mathcal{F}(N)^{2
n}}\Big{)}
\, .
\end{align}
As we can see from Eqs. (\ref{explicitpot1}) and (\ref{sder}), the
complexity of the resulting mimetic potential and Lagrange multiplier
increases, when the $F(R)$ gravity becomes more involved. In the next
section we investigate the existence of de Sitter points in mimetic $F(R)$
gravity and study their stability, when these exist.

\section{Existence and Stability of de Sitter Solutions}

In this section we address in detail the issue of existence of de Sitter
solutions and their stability within the theoretical framework of mimetic
$F(R)$ gravity with potential and Lagrange multiplier.
(Note that general condition of de Sitter solution existence in $F(R)$
gravity was derived in \cite{cognola}.)
 The acceleration eras are described by de Sitter or quasi de Sitter
solutions, so it is important to investigate the existence of de Sitter
solutions. Also the dynamical stability of the de Sitter solutions against
linear perturbations is very important, since the instability can cause
curvature perturbations which may provide a mechanism for the graceful exit
from inflation \cite{linde1,sergeitracepaper}. Another way to interpret the
dynamical instability of de Sitter solutions against linear perturbations is
that the instability indicates that the solution is not the final attractor
of the theory and thereby the eternal inflation is avoided
\cite{linde1,sergeitracepaper}.

Let us first investigate the case of the $F(R)$ gravity given in Eq.
(\ref{staro}). Consider a de Sitter solution of the form, $H(t)=H_{dS}$, for
which, the mimetic potential $V(t)$ and the Lagrange multiplier $\lambda
(t)$ become equal to,
\begin{equation}\label{avtadesitte}
V(t)=6 H_{dS}^2+144 f H_{dS}^4-1728 d H_{dS}^6,\,\,\, \lambda (t)=-3
H_{dS}^2-432 d H_{dS}^6\, .
\end{equation}
Using these and upon substitution to Eq. (\ref{diffeqnnewaddition}), we
obtain the following de Sitter solution,
\begin{equation}\label{equationsol}
H_{dS}=\frac{1}{6} \sqrt{\frac{f}{d}+\frac{\sqrt{3 d+f^2}}{d}} \, .
\end{equation}
In order to study whether this de Sitter solution is stable or not, we
consider linear perturbations of this solution of the following form,
\begin{equation}\label{perturbation}
H(t)=H_{dS}+\Delta H(t)\, ,
\end{equation}
assuming of course that the perturbation function $\Delta H (t)$ is very
small, that is, $\mid\Delta H(t)\mid\ll 1$. Substituting Eq.
(\ref{perturbation}) in Eq. (\ref{diffeqnnewaddition}) and by keeping terms
linear to $\Delta H(t)$ and to the higher derivatives $\Delta \dot{H}(t)$
and $\Delta \ddot{H}(t)$, we obtain the differential equation satisfied by
the linear perturbation $\Delta H(t)$, which is,
\begin{align}\label{ddotdiffeqna}
& -6 H_{dS}^2-144 d H_{dS}^4+2592 d H_{dS}^6-12 H_{dS} \Delta H (t)-576 d
H_{dS}^3 \Delta H (t)+15552 d H_{dS}^5 \Delta H (t)-6 \Delta\dot{ H} (t)\\
\notag &-216 d H_{dS}^2 \Delta\dot{ H} (t)+288 d H_{dS}^3 \Delta\dot{ H}
(t)+5184 d H_{dS}^4 \Delta\dot{ H} (t)\\ \notag & -10368 d H_{dS}^5
\Delta\dot{ H} (t)+72 d H_{dS}^2 \Delta\ddot{ H} (t)-2592 d H_{dS}^4
\Delta\ddot{ H} (t)=0\, ,
\end{align}
with the ``dot'' denoting differentiation with respect to the cosmic time.
The differential equation (\ref{ddotdiffeqna}) can analytically be solved,
with the solution being,
\begin{equation}\label{solutionbigfr3}
\Delta H(t)=\mathcal{A}+c_Ae^{\delta_1 t}+c_Be^{\delta_2 t}\, ,
\end{equation}
with $c_A$ and $c_B$ integration constants and the parameters $\mathcal{A}$,
$\delta_1$ and $\delta_2$ appear in the Appendix A. Substituting the de
Sitter value $H_{dS}$ we found in Eq. (\ref{equationsol}) in Eq.
(\ref{solutionbigfr3}), we observe that the parameter $\delta_2$ is positive
while the parameter $\delta_1$ is negative, and hence the term $e^{\delta_2
t}$ dominates the evolution of the linear perturbation $\Delta H(t)$, hence
the solution (\ref{solutionbigfr3}) is unstable which means that it is not
the final attractor of the theory. Therefore, a graceful exit from inflation
in achieved for the model (\ref{staro}), in the context of mimetic $F(R)$
gravity.

In a similar way, let us investigate the de Sitter solutions of a quite well
known class of viable $F(R)$ cosmological models, that of Eq.
(\ref{generalfrmodels}). Particularly, consider the following model,
\begin{equation}\label{generalfrmodelsnew}
F(R)=R-\mu R^n-d R^{-p}\, ,
\end{equation}
where $\mu ,n,d,p$ are positive arbitrary real numbers. In the large
curvature limit, the $F(R)$ gravity of Eq. (\ref{generalfrmodelsnew})
becomes,
\begin{equation}\label{newmodelfr}
F(R)\simeq R-\mu  R^n \, ,
\end{equation}
where the parameter $\mu$ is a positive phenomenological parameter. Note
that when $n$ takes values in the interval $0<n<1$, the model of Eq.
(\ref{newmodelfr}) is a quite well known phenomenological model
\cite{reviews1}, but we will assume that $n$ takes positive arbitrary
values.
 The mimetic potential $V(t)$ and the Lagrange
 multiplier $\lambda (t)$, for the Hubble rate being $H(t)=H_{dS}$, become
equal to,
\begin{equation}\label{avtadesitte12}
V(t)=6 HdS^2 - 12^n HdS^{2n} \mu,\,\,\, \lambda (t)=-3 H_{dS}^2+2^{-1+2 n}
3^n H_{dS}^{2n} \mu -3^n 4^{-1+n} H_{dS}^{2n} n \mu\, ,
\end{equation}
and upon substitution to Eq. (\ref{diffeqnnewaddition}), we obtain the only
de Sitter solution that exists for this model, which is,
\begin{equation}\label{equationsol12}
H_{dS}=\left(2^{-2+2 n} 3^{-1+n} n \mu \right)^{\frac{1}{2-2 n}} \, .
\end{equation}
As before, what now remains is to study the stability of this de Sitter
vacuum. Following the research line of the previous paradigm, we obtain the
following differential equation satisfied by the linear perturbation $\Delta
H(t)$,
\begin{align}\label{ddotdiffeqna12}
& -6 H_{dS}^2-12 H_{dS} \Delta H (t)-6 \Delta\dot{ H} (t)+144 H_{dS}^3 n \mu
\Delta\dot{ H} (t) \left(12 (H_{dS})^2\right)^{-2+n}\\ \notag & -144
H_{dS}^3 n^2 \mu  \Delta\dot{ H} (t) \left(12 (H_{dS})^2\right)^{-2+n} +6
H_{dS}^2 n \mu \left(12 (H_{dS})^2\right)^{-1+n}+12 H_{dS} n \mu  \Delta H
(t) \left(12 (H_{dS})^2\right)^{-1+n}\\ \notag &+6 n \mu  \Delta\dot{ H} (t)
\left(12 (H_{dS})^2\right)^{-1+n} +36 H_{dS}^2 n \mu  \left(12
(H_{dS})^2\right)^{-2+n} \Delta\ddot{ H} (t)\\ \notag &-36 H_{dS}^2 n^2 \mu
\left(12 (H_{dS})^2\right)^{-2+n} \Delta\ddot{ H} (t)=0\, ,
\end{align}
with the ``dot'' again denoting differentiation with respect to the cosmic
time $t$. The solution to this differential equation is,
\begin{equation}\label{solutionbigfr312}
\Delta H(t)=-\frac{H_{dS}}{2}+d_Ae^{\zeta_1 t}+d_Be^{\zeta_2 t}\, ,
\end{equation}
where $\zeta_1$ and $\zeta_2$ are given in the Appendix A, while the
parameters $d_A$ and $d_B$ are arbitrary integration constants.

By substituting the exact de Sitter value $H_{dS}$ we found in Eq.
(\ref{equationsol12}), in the parameter $\zeta_2$, it can easily proven that
when $\frac{n}{\mu}\gg 1$, the parameter $\zeta_2$ is positive while when
$\frac{n}{\mu}\ll 1$, the parameter $\zeta_2$ is negative. Therefore, when
$\frac{n}{\mu}\gg 1$, the term $\sim e^{\zeta_2 t}$ dominates the evolution,
since the other term contains $\zeta_1$, is always negative. Therefore, when
$\frac{n}{\mu}\gg 1$, the de Sitter solution we found in the model of Eq.
(\ref{newmodelfr}) is unstable. Therefore, this indicates that the de Sitter
solution is not the final attractor of the theory and therefore inflation
can end in this model, always in the presence of mimetic potential and
Lagrange multiplier.

Now consider the small curvature limit of the $F(R)$ gravity
(\ref{generalfrmodelsnew}), in which case it can be approximated by,
\begin{equation}\label{generalfrmodelsnewsmallR}
F(R)\simeq R-d R^{-p}\, .
\end{equation}
Following the same procedure as in the large curvature limit, we easily find
the following de Sitter point,
\begin{equation}\label{desittsmallcurv}
H_{dS}=\left(2^{-2+2 p} 3^{-1+p} d p\right)^{\frac{1}{2-2 p}}\, ,
\end{equation}
and by repeating the same procedure, it can be proved that the de Sitter
point (\ref{desittsmallcurv}) is stable against linear perturbations. In
fact, the resulting differential equation of perturbations and solutions can
easily be obtained from Eqs. (\ref{avtadesitte12}), (\ref{ddotdiffeqna12})
and (\ref{solutionbigfr312}), by replacing $n\rightarrow -p,\mu \rightarrow
d$ Therefore we have a quite appealing case at hand, in which the $F(R)$
gravity (\ref{generalfrmodelsnew}), in the large curvature limit has the
unstable de Sitter point (\ref{equationsol12}), and in the small curvature
limit it has a stable de Sitter point (\ref{desittsmallcurv}). Therefore,
since the large curvature describes the inflationary era, it is possible
that the cosmological evolution starts from the initial, large curvature de
Sitter, and due to the fact that it is unstable, graceful exit is
guaranteed. After that, the system might end up to the late-time, small
curvature, de Sitter point, which is stable, so the system has reached a
final attractor. Hence the unification of late and early-time acceleration
is a viable possibility in this model.

Before closing this section, we briefly discuss the results corresponding to
another model of a mimetic $F(R)$ gravity, with the $F(R)$ function being,
\begin{equation}\label{exponentialmodel}
F(R)=R-\Lambda_{eff}\left (1-e^{-b R}\right)\, .
\end{equation}
This is an exponential model, and the exponential $F(R)$ models are known to
have many appealing properties \cite{exponentialmodels}. The model
(\ref{exponentialmodel}) possesses two de Sitter vacua, which are,
\begin{equation}\label{desittervac1}
H_{dS_1}=\frac{\sqrt{\ln\left[(b \Lambda )^{1/12}\right]}}{\sqrt{b}},\,\,\,
H_{dS_2}=\frac{\sqrt{\ln[b \Lambda ]}}{2 \sqrt{3} \sqrt{b}}\, ,
\end{equation}
and by using the same research line as in the previous two examples, it can
be proven that these two de Sitter vacua are stable, so graceful exit does
not occur.

In table (\ref{TableI}) we summarize the results for some of the mimetic
$F(R)$ models we studied, with regards their compatibility with the
observational data and also with respect to the exit from inflation.
\begin{table*}[h]
\small
\caption{\label{TableI}Some Mimetic $F(R)$ models confronted with
observational data and with exit from inflation}
\begin{tabular}{@{}crrrrrrrrrrr@{}}
\tableline
\tableline
\tableline
F(R) model & Compatibility with Observations  & $\,\,\,\,\,\,\,$Exit from
Inflation$/$ Stability of dS Vacua
\\\tableline
$F(R)=R-d R^3+f R^2$,  & Yes
$\,\,\,\,\,\,\,\,\,\,\,\,\,$$\,\,\,\,\,\,\,\,\,\,\,\,\,$$\,\,\,\,\,\,\,\,\,\
,\,\,\,$ & Yes when $d>0$ and $f>0$, unstable dS \\
\tableline
$F(R)=R-\mu  R^n $  &
Yes$\,\,\,\,\,\,\,\,\,\,\,\,\,$$\,\,\,\,\,\,\,\,\,\,\,\,\,$$\,\,\,\,\,\,\,\,
\,\,\,\,\,\,\,$ & Yes when $\frac{n}{\mu}\gg 1$, unstable dS\\
\tableline
$F(R)=R-d  R^{-p} $  &
Yes$\,\,\,\,\,\,\,\,\,\,\,\,\,$$\,\,\,\,\,\,\,\,\,\,\,\,\,$$\,\,\,\,\,\,\,\,
\,\,\,\,\,\,\,$ & No, stable dS\\
\tableline
$F(R)=R-\Lambda_{eff}\left (1-e^{-b R}\right)$  &
Yes$\,\,\,\,\,\,\,\,\,\,\,\,\,\,\,$$\,\,\,\,\,\,\,\,\,\,\,\,\,\,\,$$\,\,\,\,
\,\,\,\,\,\,\,$ & No, stable dS\\
\tableline
\tableline
 \end{tabular}
\end{table*}
Before closing this section an important comment is in order. For a generic inflationary cosmology, the background solution is still stable against the back-reaction of curvature perturbations, since the background solution is often a local attractor of the theory. Admittedly, with the de Sitter solutions stability analysis we performed in this section, it is not easy to see whether the instability would not
only enforce the Universe to exit the inflationary phase, but may also spoil the predictions for primordial power spectra that are important
to explain the CMB data, since indeed the curvature perturbations may grow in such a way so that the primordial spectrum is significantly affected. This issue is quite important and should be addressed concretely in a future work. This however exceeds the purposes of this paper, so in a future work we shall address it in detail.

\section{Known Cosmologies from Mimetic $F(R)$ with Potential}

With mimetic $F(R)$ theories it is quite easy to realize various
cosmological scenarios, given the Hubble rate and the $F(R)$ gravity. Then,
by using the equations of motion of the mimetic $F(R)$ theory, namely Eqs.
(\ref{scalarpot}), (\ref{lagrange}) and (\ref{diffeqnnewaddition}), we can
easily find the mimetic potential and the Lagrange multiplier that generates
the given cosmological evolution. In principle, one can use any viable
$F(R)$ gravity, so we demonstrate this method by using the model of Eq.
(\ref{newmodelfr}), and we try to realize various cosmological scenarios. We
start off with a Type IV singular evolution
\cite{Nojiri:2005sx,sergeioikonomou1,sergeioikonomou2,sergeioikonomou3,serge
ioikonomou4}, for which the Hubble rate has the following general form,
\begin{equation}\label{hubdfre}
H(t)=f_0\left ( t-t_s \right)^{\alpha}\, ,
\end{equation}
where $\alpha$ is a positive number of the following form,
\begin{equation}\label{IV2}
 \alpha= \frac{2n}{2m + 1}\, ,
\end{equation}
with $m$ and $n$ integer numbers. Also, in order for a Type IV singularity
to be realized, the parameter $\alpha$ has to be $\alpha>1$, so $n$ and $m$
must be chosen accordingly. Recall that a Type IV singularity is not a
crushing type singularity, and it is the most ''mild'' among the four types
of finite time cosmological singularities, which were classified in Ref.
\cite{Nojiri:2005sx}. For a Type IV singularity, as $t \to t_s$, the scale
factor, the effective energy density and the effective pressure are finite,
that is, $a \to a_s$, $\rho_\mathrm{eff} \to \rho_s$,
$\left|p_\mathrm{eff}\right| \to p_s$. In addition, both the Hubble rate and
it's first derivative are finite, but the second or higher
derivatives of the Hubble rate are singular, as $t \to t_s$. So by using
Eqs. (\ref{scalarpot}), (\ref{lagrange}), the $F(R)$ gravity of Eq.
(\ref{newmodelfr}) and also the relation $R=6\left (\dot{H}+2H^2 \right )$,
we easily obtain the scalar potential $V(\phi)$, which is,
\begin{align}\label{vtsingularevolution}
& V(\phi)=6 f_0 \alpha  (-t_s+\phi )^{-1+\alpha }+12 f_0^2 (-t_s+\phi )^{2
\alpha } \\ \notag & -2 \Big{(}f_0 \alpha  (-t_s+\phi )^{-1+\alpha }+3 f_0^2
(-t_s+\phi )^{2 \alpha }\Big{)}-\mu  \Big{(}6 f_0 \alpha  (-t_s+\phi
)^{-1+\alpha }+12 f_0^2 (-t_s+\phi )^{2 \alpha }\Big{)}^n \\ \notag &
-4 f_0 (-1+n) n \mu  (-t_s+\phi )^{\alpha }
\Big{(}6 f_0 \alpha  (-t_s+\phi )^{-1+\alpha }+12 f_0^2 (-t_s+\phi )^{2
\alpha }\Big{)}^{-2+n} \times \\ \notag &\Big{(}6 f_0 (-1+\alpha ) \alpha
(-t_s+\phi )^{-2+\alpha }  +24 f_0^2 \alpha  (-t_s+\phi )^{-1+2 \alpha
}\Big{)}\\ \notag &
-2 n \mu  \Big{(}(-1+n) \Big{(}6 f_0 \alpha  (-t_s+\phi )^{-1+\alpha }+12
f_0^2 (-t_s+\phi )^{2 \alpha }\Big{)}^{-2+n} \Big{(}6 f_0 (-2+\alpha )
(-1+\alpha ) \alpha  (-t_s+\phi )^{-3+\alpha } \\ \notag &+24 f_0^2 \alpha
(-1+2 \alpha ) (-t_s+\phi )^{-2+2 \alpha }\Big{)}+(-2+n) (-1+n) \times \\
\notag & \Big{(}6 f_0 \alpha  (-t_s+\phi )^{-1+\alpha }+12 f_0^2 (-t_s+\phi
)^{2 \alpha }\Big{)}^{-3+n} \Big{(}6 f_0 (-1+\alpha ) \alpha  (-t_s+\phi
)^{-2+\alpha }+24 f_0^2 \alpha  (-t_s+\phi )^{-1+2 \alpha }\Big{)}^2\Big{)}
\end{align}
and the Lagrange multiplier function $\lambda (\phi)$ is equal to,
\begin{align}\label{lagrphicosmconstr}
\lambda (\phi)&=3 f_0 (-1+n) n \mu  (-t_s+\phi )^{\alpha } \Big{(}6 f_0
\alpha (-t_s+\phi )^{-1+\alpha }+12 f_0^2 (-t_s+\phi )^{2 \alpha
}\Big{)}^{-2+n} \times \\ \notag & \Big{(}6 f_0 (-1+\alpha ) \alpha
(-t_s+\phi )^{-2+\alpha }+24 f_0^2 \alpha  (-t_s+\phi )^{-1+2 \alpha
}\Big{)}\, .
\end{align}
In the same way, we can realize other cosmological scenarios, which are
``exotic'' for the standard Einstein-Hilbert gravity. For example, the
matter bounce scenario \cite{matterbounce,matterbouncesergeioikonomou}, can
be easily realized with mimetic $F(R)$ gravity. The Hubble rate of the
matter bounce scenario is equal to,
\begin{equation}\label{mattbouncescenarohib}
H(t)=\frac{4 t \rho }{3 \left(1+t^2 \rho \right)}\, ,
\end{equation}
were $\rho $ is a parameter related to the critical density of LQC theory
(see \cite{matterbounce} for details). This scenario is known to be realized
in the context of Loop Quantum Cosmology, but it is possible to realize this
in the context of $F(R)$ gravity, as was explicitly shown in
\cite{matterbouncesergeioikonomou,oikonomou1}. However, the resulting $F(R)$
that can generate the matter bounce has a very specific form, so the
non-mimetic $F(R)$ gravity theoretical framework does not allow to use a
general, in principle viable, $F(R)$ gravity. Nevertheless, the mimetic
$F(R)$ gravity with Lagrange multiplier formalism allows to use any $F(R)$
gravity, so for simplicity, we use again the $F(R)$ gravity of Eq.
(\ref{newmodelfr}), and we investigate which potential $V(\phi )$ and
Lagrange multiplier $\lambda (\phi )$ can give rise to the matter bounce
scenario, with Hubble rate given in Eq. (\ref{mattbouncescenarohib}). As in
the previous case, by using Eqs. (\ref{scalarpot}), (\ref{lagrange}), the
$F(R)$ gravity of Eq. (\ref{newmodelfr}) and also Eq.
(\ref{mattbouncescenarohib}), we obtain the following mimetic potential
$V(\phi )$,
\begin{align}\label{mimetmattbounce}
& V(\phi )= \frac{64 \rho ^2 \phi ^2}{3 \left(1+\rho  \phi
^2\right)^2}-\frac{8 \rho  \left(-1+\rho  \phi ^2\right)}{\left(1+\rho  \phi
^2\right)^2}-\frac{8 \rho  \left(1+3 \rho  \phi ^2\right)}{3 \left(1+\rho
\phi ^2\right)^2}-\left(\frac{8}{3}\right)^n \mu  \left(\frac{\rho
\left(3+5 \rho \phi ^2\right)}{\left(1+\rho  \phi ^2\right)^2}\right)^n\\
\notag &
+\frac{2^{2+3 n} 3^{-n} (-1+n) n \mu  \rho  \phi ^2 \left(1+5 \rho  \phi
^2\right) \left(\frac{\rho  \left(3+5 \rho  \phi ^2\right)}{\left(1+\rho
\phi ^2\right)^2}\right)^n}{\left(3+5 \rho  \phi ^2\right)^2}\\ \notag &
-\frac{2^{-1+3 n} 3^{1-n} (-1+n) n \mu  \left(\frac{\rho  \left(3+5 \rho
\phi ^2\right)}{\left(1+\rho  \phi ^2\right)^2}\right)^n \left(-3+(-39+2 n)
\rho \phi ^2+5 (-9+4 n) \rho ^2 \phi ^4+25 (-1+2 n) \rho ^3 \phi
^6\right)}{\left(3+5 \rho  \phi ^2\right)^3}\, ,
\end{align}
and the corresponding Lagrange multiplier $\lambda (\phi)$ is equal to,
\begin{align}\label{lagrancematterbounce}
& \lambda (\phi)=\frac{64 \rho ^2 \phi ^2}{6 \left(1+\rho  \phi
^2\right)^2}+\left(\frac{8}{3}\right)^n \frac{\mu }{2} \left(\frac{\rho
\left(3+5 \rho  \phi ^2\right)}{\left(1+\rho  \phi ^2\right)^2}\right)^n
-\frac{2^{1+3 n} 3^{1-n} (-1+n) n \mu  \rho  \phi ^2 \left(1+5 \rho  \phi
^2\right) \left(\frac{\rho  \left(3+5 \rho  \phi ^2\right)}{\left(1+\rho
\phi ^2\right)^2}\right)^n}{2\left(3+5 \rho  \phi ^2\right)^2}\\ \notag &
+\frac{8 \rho  \left(3+\rho  \phi ^2\right)
\left(1-\left(\frac{3}{8}\right)^{1-n} n \mu  \left(\frac{\rho  \left(3+5
\rho \phi ^2\right)}{\left(1+\rho  \phi ^2\right)^2}\right)^{-1+n}\right)}{6
\left(1+\rho  \phi ^2\right)^2}+\frac{8 \rho  \left(-1+\rho  \phi
^2\right)}{2\left(1+\rho  \phi ^2\right)^2}\, .
\end{align}
As we can see, different mimetic potentials and Lagrange multipliers can
give rise to different cosmological scenarios, by using the same $F(R)$
gravity. As a final example we shall present the mimetic potential $V(\phi)$
and the Lagrange multiplier $\lambda (\phi)$, which can give rise to the
superbounce scenario \cite{superbounce}, with scale factor and Hubble rate
given below,
\begin{equation}\label{superbouncescenariob}
a(t)=\left(t-t_s \right)^{\frac{2}{c^2}}, \,\,\, H(t)=\frac{2}{c^2
(t-t_s)}\, ,
\end{equation}
with $c$ a parameter determined by the theory, see \cite{superbounce} for
details. Following the steps of the previous sections, we obtain the
following potential $V(\phi )$,
\begin{align}\label{finalsuperbounce}
V(\phi )=\frac{3^{-1+n} 4^n \left(12+c^2 \left(-3+4 n-4 n^2\right)+c^4 n
\left(1-3 n+2 n^2\right)\right) \mu  \left(-\frac{-4+c^2}{c^4 (t_s-\phi
)^2}\right)^n}{-4+c^2}-\frac{8 \left(-3+c^2\right)}{c^4 (t_s-\phi )^2}
\end{align}
and the corresponding Lagrange multiplier $\lambda (\phi )$ is equal to,
\begin{align}\label{lagrasuperbounce}
& \lambda (\phi )= 2^{-1+2 n} 3^n \mu  \left(-\frac{-4+c^2}{c^4 (t_s-\phi
)^2}\right)^n+\frac{12^n c^2 (-1+n) n \mu  \left(-\frac{-4+c^2}{c^4
(t_s-\phi )^2}\right)^n}{-4+c^2} \\\ \notag &
+\frac{6 \left(-4+c^2\right)}{c^4 (t_s-\phi )^2}-\frac{6 \left(-2+c^2\right)
\left(1-12^{-1+n} n \mu  \left(\frac{4-c^2}{c^4 (t_s-\phi
)^2}\right)^{-1+n}\right)}{c^4 (t_s-\phi )^2}\, .
\end{align}
The inverse procedure is possible as we show in a later section.
Particularly, given the mimetic potential $V(t=\phi)$ and the Lagrange
multiplier $\lambda (t=\phi)$, it is possible to find the $F(R)$ gravity
which generates an also given cosmological evolution. In principle the
mimetic $F(R)$ gravity that realizes a specified cosmological evolution, is
different from the pure $F(R)$ gravity that can realize the same
cosmological evolution.

\section{Einstein Frame Counterpart of Mimetic $F(R)$ gravity with
Potential}

Having studied in detail all the Jordan frame implications of the mimetic
$F(R)$ gravity, it is worth discussing the Einstein frame counterpart
theory, which is a two scalar field theory. The multi-scalar field theory is
motivated mainly from the area of high-energy physics, in which multi-scalar
field theories play a prominent role, especially in relation with inflation
\cite{9807278}. One characteristic feature of multi-field models for
inflation is that these models produce isocurvature perturbations, which in
effect can generate non-Gaussianities, when long-wave length modes are
considered \cite{kaizer}. In this section we shall be interested in studying
the generic properties of the Einstein frame counterpart theory of the
Jordan frame mimetic $F(R)$ theory and we shall also calculate the slow-roll
indices of the resulting two-scalar theory. We shall also exemplify our
findings by studying a theory which in the Jordan frame is the $R^2$ theory.

We start off by producing the Einstein frame counterpart action of the
action given in Eq. (\ref{actionmimeticfraction}), which can be done
following standard procedures (see for example \cite{reviews1}) in the
following way: We introduce, the auxiliary scalar field $A$, and the action
of Eq. (\ref{actionmimeticfraction}), becomes,

\begin{align}\label{actionmimeticfractioneinsteinfr}
& S=\int \mathrm{d}x^4\sqrt{-g}\Big{ (} \left ( F'(A)(R-A)+F(A) \right
)-V(\phi)+\lambda \left(g^{\mu \nu}\partial_{\mu}\phi\partial_{\nu}\phi
+1\right)\Big{ )}
\end{align}
By varying the action of Eq. (\ref{actionmimeticfractioneinsteinfr}) with
respect to the auxiliary scalar field $A$, we obtain the solution $A=R$,
which implies the mathematical equivalence of the two actions  given in Eqs.
(\ref{actionmimeticfraction}) and (\ref{actionmimeticfractioneinsteinfr}).
We can easily obtain the Einstein frame counterpart theory of
(\ref{actionmimeticfractioneinsteinfr}), by performing the following
conformal transformation,
\begin{equation}\label{can}
\varphi =-\sqrt{\frac{3}{2\kappa^2}}\ln (F'(A))
\end{equation}
where the scalar field $\varphi$ is the Einstein frame inflaton field. Note
that in the theory there exists another scalar field $\phi$, from the
mimetic theory.
In addition, the Jordan frame metric $g_{\mu \nu}$ is transformed to a new
metric in the Einstein frame, which we denote as $\hat{g}_{\mu \nu }$. These
two metric are related as follows,
\begin{equation}\label{conftransmetr}
g_{\mu \nu}=e^{-\frac{2}{3}\kappa \varphi }\hat{g}_{\mu \nu }
\end{equation}
Moreover, the expression $\sqrt{-g}$ is transformed as $\sqrt{-g}=e^{-2
\sqrt{\frac{2}{3}}\kappa \varphi }\sqrt{-\hat{g}}$, so the resulting
Einstein frame action reads,
\begin{align}\label{einsteinframeaction}
& \mathcal{\tilde{S}}=\int \mathrm{d}^4x\sqrt{-\hat{g}}\Big{ (}
\frac{\hat{R}}{2\kappa^2}-\frac{1}{2}\left (\frac{F''(A)}{F'(A)}\right
)^2\hat{g}^{\mu \nu }\partial_{\mu }A\partial_{\nu }A
-\frac{1}{2\kappa^2}\left ( \frac{A}{F'(A)}-\frac{F(A)}{F'(A)^2}\right )  \\
\notag &
-V(\phi)e^{-2\sqrt{\frac{2}{3}}\kappa \varphi }+\lambda (\phi) \hat{g}^{\mu
\nu}e^{-\sqrt{\frac{2}{3}}\kappa \varphi
}\partial_{\mu}\phi\partial_{\nu}\phi+\lambda
(\phi)e^{-2\sqrt{\frac{2}{3}}\kappa \varphi }\Big{)}\, ,
\end{align}
which can be written as follows,
\begin{align}\label{einsteinframeaction1}
& \mathcal{\tilde{S}}=\int \mathrm{d}^4x\sqrt{-\hat{g}}\Big{ (}
\frac{\hat{R}}{2\kappa^2}-\frac{1}{2}\hat{g}^{\mu \nu }\partial_{\mu
}\varphi\partial_{\nu }\varphi -U(\varphi )
-V(\phi)e^{-2\sqrt{\frac{2}{3}}\kappa \varphi }+\lambda (\phi) \hat{g}^{\mu
\nu}e^{-\sqrt{\frac{2}{3}}\kappa \varphi
}\partial_{\mu}\phi\partial_{\nu}\phi+\lambda
(\phi)e^{-2\sqrt{\frac{2}{3}}\kappa \varphi }\Big{)}\, ,
\end{align}
where the exact form of the potential $U(\varphi )$, as a function of the
scalar field $\varphi $, is equal to,
\begin{align}\label{potentialvsigma}
U(\varphi )=\frac{A}{F'(A)}-\frac{F(A)}{F'(A)^2}=\frac{1}{2\kappa^2}\left (
e^{\sqrt{2\kappa^2/3}\varphi }\hat{R}\left (e^{-\sqrt{2\kappa^2/3}\varphi}
\right )- e^{2\sqrt{2\kappa^2/3}\varphi }F\left [ \hat{R}\left
(e^{-\sqrt{2\kappa^2/3}\varphi} \right ) \right ]\right )\, .
\end{align}
For later convenience, we introduce the potential $\tilde{V}(\varphi,\phi)$,
which is equal to,
\begin{equation}\label{vtilde}
\tilde{V}(\varphi,\phi)=-U(\varphi )
-V(\phi)e^{-2\sqrt{\frac{2}{3}}\kappa \varphi }+\lambda
(\phi)e^{-2\sqrt{\frac{2}{3}}\kappa \varphi }\, ,
\end{equation}
and by using this, the action of Eq. (\ref{einsteinframeaction1}), is
written as follows,
\begin{equation}\label{einsteinframea}
\mathcal{\tilde{S}}=\int \mathrm{d}^4x\sqrt{-\hat{g}}\Big{ (}
\frac{\hat{R}}{2\kappa^2}-\frac{1}{2}\hat{g}^{\mu \nu }\partial_{\mu
}\varphi\partial_{\nu }\varphi+\lambda (\phi) \hat{g}^{\mu
\nu}e^{-\sqrt{\frac{2}{3}}\kappa \varphi
}\partial_{\mu}\phi\partial_{\nu}\phi-\tilde{V}(\varphi,\phi)\Big{)}\, ,
\end{equation}
The interesting of the quite general Lagrangian of Eq.
(\ref{einsteinframeaction1}), is that it resembles some phenomenologically
interesting models, like the ones appearing in \cite{linde} and
\cite{nanopoulos}. The models studied in \cite{linde}, are imaginary
deformations of the Starobinsky model \cite{starobinskyold}, and the
Lagrangian of these models is,
\begin{equation}\label{imaginarystarlagrangian}
\mathcal{L}=\frac{1}{2}R-\frac{1}{2}\partial_{\mu }\varphi\partial_{\nu
}\varphi-\frac{1}{2}e^{-2\sqrt{\frac{2}{3}}\varphi}\partial_{\mu
}\phi\partial_{\nu
}\phi-\frac{1}{2}M^2e^{-2\sqrt{\frac{2}{3}}\varphi}\phi^2-\frac{3}{4}M^2\left( 1-e^{-\sqrt{\frac{2}{3}}\varphi}\right)^2\, .
\end{equation}
This model is related to minimal supergravity approaches \cite{linde} and it
is similar to the one studied in \cite{nanopoulos}. Obviously, if we choose
the potential $U(\varphi)$ in Eq. (\ref{einsteinframeaction1}), to be the
Starobinsky potential \cite{starobinskyold}, that is,
\begin{equation}\label{starobpot}
U(\varphi )=C_0\left( 1-e^{-\sqrt{\frac{2}{3}}\kappa \varphi}\right)^2\, ,
\end{equation}
where $C_0=\frac{3}{4}\kappa^2 M^2$, the resemblance between the two scalar
models of Eq. (\ref{einsteinframeaction1}) and
(\ref{imaginarystarlagrangian}) is obvious. Note that the corresponding
Jordan frame theory would be,
\begin{equation}\label{actionmimeticfractionjordanframestaro}
S=\int \mathrm{d}x^4\sqrt{-g}\left ( R+\alpha R^2-V(\phi)+\lambda
\left(g^{\mu \nu}\partial_{\mu}\phi\partial_{\nu}\phi +1\right)\right )\, ,
\end{equation}
where $\alpha=\frac{1}{8\kappa^2 C_0}$. Actually, the Lagrangian of Eq.
(\ref{einsteinframeaction1}) can serve as a generalization of the model
(\ref{imaginarystarlagrangian}). What interests us is to calculate the
observational indices of inflation for the two scalar model and we shall use
the techniques developed in \cite{kaizer,twoscalarindices}. In the next
section, we calculate the slow-roll indices in the general case, and we
exemplify our findings in the case the potential $U(\varphi)$ is chosen as
in Eq. (\ref{starobpot}).

\subsection{Double Scalar Model Calculation of Observational Indices}

In this section, we shall calculate the slow-roll indices for the two scalar
model of Eq. (\ref{einsteinframeaction1}) and we exemplify our findings for
the case that the scalar potential $U(\varphi)$ is the Starobinsky potential
of Eq. (\ref{starobpot}). We shall use two different approaches for this
calculation, with the first being the one developed in \cite{kaizer}, and
the second one being the one used in \cite{barrowslowroll}. With regards to
the latter, we directly compute the Hubble slow-roll parameters, without
assuming any slow-roll for the fields. We have to note that at first order
in the slow-roll expansion, the two approaches are equivalent, at least when
the first slow-roll index takes small values.

We shall use the formalism and notation of \cite{kaizer}. Consider the
following multi-scalar field action,
\begin{equation}\label{miltuiscale}
S=\int \mathrm{d}^4x\sqrt{-\hat{g}}\Big{ (}
\frac{\hat{R}}{2\kappa^2}-\frac{1}{2}G_{IJ}(\phi^I)\hat{g}^{\mu \nu
}\partial_{\mu }\phi^I\partial_{\nu }\phi^J -V(\phi^{I} ) \Big{)}\, ,
\end{equation}
with $I,J=1,2$ and in our case the various scalar fields $\phi^I$ are
$\phi^1=\varphi$ and $\phi^2=\phi$. In addition, the metric $G_{IJ}(\phi^I)$
is scalar field dependent and it corresponds to the metric in the scalar
field configuration space, which in our case is two dimensional.
Specifically, the matrix representation of the metric in the case the action
is as in (\ref{einsteinframea}), is equal to,
\begin{equation}\label{metricconfspace}
G=\left(
\begin{array}{cc}
 1 & 0 \\
 0 & e^{-\sqrt{6} \kappa  \varphi } \lambda (\phi ) \\
\end{array}
\right)\, .
\end{equation}
The equations of motion corresponding to the action (\ref{miltuiscale}) are
equal to,
\begin{align}\label{gfgdgd}
& H^2=\frac{\kappa^2}{3}\left(
\frac{1}{2}G_{IJ}\dot{\varphi}^I\dot{\varphi}^J+V(\varphi^I)\right)\\ \notag
&
\dot{H}=-\frac{1}{\kappa^2}G_{IJ}\dot{\varphi}^I\dot{\varphi}^J\\ \notag &
\square \phi^I+\hat{g}^{\mu \nu }\Gamma^I_{IK}\partial_{\mu
}\phi^J\partial_{\nu }\phi^K-G^{IK}V_{,K}=0\, ,
\end{align}
where $V_{,K}=\partial V/\partial \phi^K$, and
$\Gamma^I_{IK}=\Gamma^I_{IK}(\varphi,\phi)$ are the Christoffel symbols for
the two dimensional field configuration space, with metric
(\ref{metricconfspace}). In the case of the action (\ref{einsteinframea}),
the non-zero Christoffel symbols read,
\begin{align}\label{christofellsymbols}
& \Gamma^{\varphi}_{\phi \phi}= \sqrt{\frac{3}{2}} e^{-\sqrt{6} \kappa
\varphi } \kappa  \lambda (\phi )  ,\,\,\,\Gamma^{\phi}_{\phi
\varphi}=-\sqrt{\frac{3}{2}} \kappa ,\,\,\, \\\notag &
\Gamma^{\phi}_{\phi \phi}=\frac{\lambda '(\phi )}{2 \lambda (\phi )},\,\,\,
\Gamma^{\phi}_{\varphi \phi }= -\sqrt{\frac{3}{2}} \kappa ,\,
\end{align}
and in this case, the prime denotes differentiation with respect to the
corresponding argument, which is $\phi$. For notational simplicity we
introduce the field $\sigma$, which is
$\dot{\sigma}=\sqrt{G_{IJ}\dot{\phi}^{I}\dot{\phi}^{J}}$, and also the
vector field $\hat{\sigma}^{I}=\frac{\dot{\phi}^I}{\dot{\sigma}}$. In terms
of these parameters, the field equations (\ref{gfgdgd}) are written as
follows,
\begin{align}\label{fieldhf}
& H^2=\frac{\kappa^2}{3}\left(\frac{1}{2}\dot{\sigma}^2+V \right), \\ \notag
&
\dot{H}=-\frac{\kappa^2}{2}\dot{\sigma}^2,\\ \notag &
\ddot{\sigma}+3H\dot{\sigma}+V_{,\sigma}=0
\end{align}
where $V_{,\sigma}=\hat{\sigma}^IV_{,I}$. In our case, the potential $V$
appearing in Eqs. (\ref{gfgdgd}) and (\ref{fieldhf}) is identified with the
potential $\tilde{V}(\varphi,\phi)$ of Eq. (\ref{einsteinframea}), so we
identify from now on $V$ with $\tilde{V}$. For convenience we compute
$\dot{\sigma}$ and the various $\hat{\sigma}^I$, which read,
\begin{align}\label{dhfhf}
& \dot{\sigma}=\sqrt{\dot{\varphi}^2+e^{-\sqrt{6} \kappa  \varphi } \lambda
(\phi )\dot{\phi}^2},\,\,\,
\hat{\sigma}^1=\frac{\dot{\varphi}}{\sqrt{\dot{\varphi}^2+e^{-\sqrt{6}
\kappa \varphi } \lambda (\phi )\dot{\phi}^2}} , \,\,\,
\hat{\sigma}^2=\frac{\dot{\phi}}{\sqrt{\dot{\varphi}^2+e^{-\sqrt{6} \kappa
\varphi } \lambda (\phi )\dot{\phi}^2}}
\end{align}
Until this point, the general framework is the same and will hold true for
every approach we adopt for the computation of the observational indices.
However, we differentiate our analysis at this point, and we adopt two
different approaches for the observational indices, the potential slow-roll
approach (with regards to the slow-roll index $\eta$), and the Hubble
slow-roll approach. For details on these issues, see \cite{barrowslowroll}.
In our case, the two approaches differ only on the second slow-roll
parameter $\eta$, so it is a hybrid version of the two aforementioned cases.

\subsubsection{Potential Related Slow-Roll Parameters}

A quite general relation that gives the first slow-roll parameter
$\epsilon$, is the following,
\begin{equation}\label{firstslowroll}
\epsilon
=-\frac{\dot{H}}{H^2}=\frac{3\dot{\sigma}^2}{\dot{\sigma}^2+\tilde{V}}\, ,
\end{equation}
which can be easily calculated for the present case. Recall that the
potential $\tilde{V}$ is given in Eq. (\ref{vtilde}), so the parameter
$\epsilon$ is equal to,
\begin{equation}\label{sgdgd}
\epsilon= \frac{3\left(\dot{\varphi}^2+e^{-\sqrt{6} \kappa  \varphi }
\lambda (\phi )\dot{\phi}^2\right)}{\dot{\varphi}^2+e^{-\sqrt{6} \kappa
\varphi } \lambda (\phi )\dot{\phi}^2+-U(\varphi )
-V(\phi)e^{-2\sqrt{\frac{2}{3}}\kappa \varphi }+\lambda
(\phi)e^{-2\sqrt{\frac{2}{3}}\kappa \varphi }}\, .
\end{equation}
Following \cite{kaizer}, the second slow-roll index $\eta$ is equal to,
\begin{equation}\label{etaslowfhdgd}
\eta=\frac{M_{\sigma \sigma}}{\kappa^2\tilde {V}}\, ,
\end{equation}
with $M_{\sigma \sigma}$ being equal to $M_{\sigma
\sigma}=\hat{\sigma}^K\hat{\sigma}^J\left(\mathcal{D}_K\mathcal{D}_J\tilde{V
}\right)$, where $\mathcal{D}_K$ stands for the covariant derivative. Note
that the covariant derivative in the field configuration space acts as
follows,
\begin{equation}\label{doublederivative}
D_JA^I=\partial_JA^I+\Gamma_{JK}^I A^K,\,\,\,
D_JA_I=\partial_JA_I-\Gamma_{JI}^K A_K\, ,
\end{equation}
Then, it easily follows that the parameter $M_{\sigma \sigma}$ can be
written,
\begin{equation}\label{etaslowfhdgd12}
M_{\sigma
\sigma}=\frac{\phi^I\phi^J}{\dot{\sigma}^2}\tilde{V}_{,KJ}-\Gamma_{KJ}^I\tilde{V}_{,I}\, .
\end{equation}
So by using the form of the potential $\tilde{V}$ of Eq. (\ref{vtilde}) and
the values of the Christoffel symbols (\ref{christofellsymbols}), we get,
\begin{align}\label{msssdetailed}
& M_{\sigma \sigma}=-6 \frac{\dot{\varphi}^2}{\dot{\sigma}^2} e^{-\sqrt{6}
\kappa  \varphi } \kappa ^2 V(\phi )+6
\frac{\dot{\varphi}^2}{\dot{\sigma}^2} e^{-\sqrt{6} \kappa  \varphi } \kappa
^2 \lambda (\phi )-3 \frac{\dot{\varphi}^2}{\dot{\sigma}^2} e^{-2 \sqrt{6}
\kappa  \varphi } \kappa ^2 V(\phi ) \lambda (\phi )-6
\frac{\dot{\varphi}\dot{\phi}}{\dot{\sigma}^2} e^{-2 \sqrt{6} \kappa
\varphi } \kappa ^2 V(\phi ) \lambda (\phi ) \\ \notag &
-3 e^{-2 \sqrt{6} \kappa  \varphi } \frac{\dot{\phi}^2}{\dot{\sigma}^2}
\kappa ^2 V(\phi ) \lambda (\phi )+3 \frac{\dot{\varphi}^2}{\dot{\sigma}^2}
e^{-2 \sqrt{6} \kappa  \varphi } \kappa ^2 \lambda (\phi )^2+6
\frac{\dot{\varphi}\dot{\phi}}{\dot{\sigma}^2} e^{-2 \sqrt{6} \kappa
\varphi } \kappa ^2 \lambda (\phi )^2+3 e^{-2 \sqrt{6} \kappa  \varphi }
\frac{\dot{\phi}^2}{\dot{\sigma}^2}  \kappa ^2 \lambda (\phi )^2\\ \notag
&-\sqrt{\frac{3}{2}} \frac{\dot{\varphi}^2}{\dot{\sigma}^2} e^{-\sqrt{6}
\kappa \varphi } \kappa  \lambda (\phi ) U'(\varphi )
-\sqrt{6} \frac{\dot{\varphi}\dot{\phi}}{\dot{\sigma}^2} e^{-\sqrt{6} \kappa
\varphi } \kappa  \lambda (\phi ) U'(\varphi )-\sqrt{\frac{3}{2}}
e^{-\sqrt{6} \kappa  \varphi } \frac{\dot{\phi}^2}{\dot{\sigma}^2}  \kappa
\lambda (\phi ) U'(\varphi )+\sqrt{\frac{3}{2}}
\frac{\dot{\varphi}^2}{\dot{\sigma}^2} e^{-\sqrt{6} \kappa  \varphi } \kappa
V'(\phi )\\ \notag & +3 \sqrt{6}
\frac{\dot{\varphi}\dot{\phi}}{\dot{\sigma}^2} e^{-\sqrt{6} \kappa  \varphi
} \kappa  V'(\phi )-\sqrt{\frac{3}{2}}
\frac{\dot{\varphi}^2}{\dot{\sigma}^2} e^{-\sqrt{6} \kappa  \varphi } \kappa
\lambda '(\phi )-3 \sqrt{6} \frac{\dot{\varphi}\dot{\phi}}{\dot{\sigma}^2}
e^{-\sqrt{6} \kappa  \varphi } \kappa  \lambda '(\phi )+\frac{e^{-\sqrt{6}
\kappa  \varphi } \frac{\dot{\phi}^2}{\dot{\sigma}^2}  V'(\phi ) \lambda
'(\phi )}{2 \lambda (\phi )}-\frac{e^{-\sqrt{6} \kappa  \varphi }
\frac{\dot{\phi}^2}{\dot{\sigma}^2}  \lambda '(\phi )^2}{2 \lambda (\phi
)}\\ \notag & +\frac{\dot{\varphi}^2}{\dot{\sigma}^2} U''(\varphi
)-e^{-\sqrt{6} \kappa  \varphi } \frac{\dot{\phi}^2}{\dot{\sigma}^2}
V''(\phi )+e^{-\sqrt{6} \kappa  \varphi }
\frac{\dot{\phi}^2}{\dot{\sigma}^2}  \lambda ''(\phi )\, .
\end{align}
We need to note that in the above equation, the prime in the case it refers
to a function of $\phi$, it obviously denotes differentiation with respect
to $\phi$, while when it refers to  a function of $\varphi$ (like in the
case of the function $U(\varphi)$), it refers to differentiation with
respect to $\varphi$. With the parameter $M_{\sigma \sigma}$, we can easily
compute the slow-roll parameter $\eta$, by simply using Eqs. (\ref{vtilde})
and (\ref{etaslowfhdgd}). In addition, in the Appendix B, we have computed
the parameter $M_{\sigma \sigma}$ for the case that the potential $U(\varphi
)$ is equal to the Starobinsky potential (\ref{starobpot}).

Before closing this section, it is worth discussing the case for which the
parameter $\epsilon$ takes small values. We could loosely state that this is
some sort of slow-roll approximation, since the higher order terms in
the slow-roll expansion of Ref \cite{barrowslowroll} can be disregarded. In
this case, the slow-roll parameter $\eta$ is approximated by \cite{kaizer},
\begin{equation}\label{etaslowrollahsgdg}
\eta \simeq \epsilon
-\frac{\ddot{\sigma}}{H\dot{\sigma}}+\mathcal{O}(\epsilon^2)\, .
\end{equation}
From the equation of motion of the $\sigma$ field appearing in Eq.
(\ref{fieldhf}), we easily obtain that,
\begin{equation}\label{voithitiki}
\ddot{\sigma}=-3H\dot{\sigma}-\hat{\sigma}^IV_{,I}\, ,
\end{equation}
which in our case is equal to,
\begin{align}\label{voithia2}
& \ddot{\sigma}=-3 H\left( \sqrt{1+e^{-\sqrt{6} \kappa  \varphi } \lambda
(\phi )} \right)-\frac{\dot{\varphi}}{1+e^{-\sqrt{6} \kappa  \varphi }
\lambda (\phi )}\Big{(}\sqrt{6} e^{-\sqrt{6} \kappa  \varphi } \kappa
V(\phi )-\sqrt{6} e^{-\sqrt{6} \kappa  \varphi } \kappa  \lambda (\phi
)+U'(\varphi )\Big{)}\\ \notag &
-\frac{\dot{\phi}}{1+e^{-\sqrt{6} \kappa  \varphi } \lambda (\phi
)}\Big{(}-e^{-\sqrt{6} \kappa  \varphi } V'(\phi )+e^{-\sqrt{6} \kappa
\varphi } \lambda '(\phi )\Big{)}\, .
\end{align}
So finally, the parameter $\eta$ in this approximation reads,
\begin{align}\label{parmeetapapaoeapproxima}
& \eta= \epsilon- \frac{-3 H\left( \sqrt{1+e^{-\sqrt{6} \kappa  \varphi }
\lambda (\phi )} \right)-\frac{\dot{\varphi}}{1+e^{-\sqrt{6} \kappa  \varphi
} \lambda (\phi )}\Big{(}\sqrt{6} e^{-\sqrt{6} \kappa  \varphi } \kappa
V(\phi )-\sqrt{6} e^{-\sqrt{6} \kappa  \varphi } \kappa  \lambda (\phi
)+U'(\varphi )\Big{)}}{H\sqrt{1+e^{-\sqrt{6} \kappa  \varphi } \lambda (\phi
)}}\\ \notag &
-\frac{-\frac{\dot{\phi}}{1+e^{-\sqrt{6} \kappa  \varphi } \lambda (\phi
)}\Big{(}-e^{-\sqrt{6} \kappa  \varphi } V'(\phi )+e^{-\sqrt{6} \kappa
\varphi } \lambda '(\phi )\Big{)}}{H\sqrt{1+e^{-\sqrt{6} \kappa  \varphi }
\lambda (\phi )}}\, .
\end{align}
In both cases, the spectral index of primordial curvature perturbations, and
the scalar-to-tensor ratio $r$, are equal to,
\begin{equation}\label{hdhdfggjuishuii}
n_s=1-6\epsilon+2\eta,\,\,\, r=16 \epsilon\, ,
\end{equation}
and as we see later, the potential slow-roll parameters are related to the
Hubble slow-roll parameters. In the next section we treat the Hubble
slow-roll parameters and the corresponding indexes.

An important remark is in order. A calculation of the indexes may reveal the
same problems that the multi-scalar inflation theories have, when these are
confronted with observations. In fact, in the present case, the potential
and the Lagrange multiplier are restricted in the Jordan frame by Eq.
(\ref{diffeqnnewaddition}) and therefore the solutions space in the
$(\varphi,\phi)$ is severely constrained. However, an interesting case is if
a trajectory with $\dot{\phi}\simeq 0$, $\phi\simeq 0$ is chosen, which
corresponds to a slightly deformed single field case. Then, the single field
approaches attributes remain in the theory, with the extra but small
non-Gaussianities caused by the second scalar field. This would be
interesting to study but exceeds the purposes of this article. Therefore,
the Jordan frame mimetic $F(R)$ gravity looks more appealing. However, for
completeness we proceed to the Hubble rate approach of the slow-roll
indices.

\subsubsection{Hubble Rate Related Slow-Roll Parameters}

In this section, we shall attempt a different approach in comparison to the
one we used in the previous section. Particularly, the difference will be on
the second slow-roll parameter $\eta $, since the first slow-roll parameter
is given by the expression in Eq. (\ref{firstslowroll}). Actually, as was
proven in Ref. \cite{barrowslowroll}, the Hubble slow-roll parameter and the
potential slow-roll parameter coincide at first order in the slow-roll
expansion. However, the second Hubble slow-roll parameter is equal to,
\begin{equation}\label{gfhsgs}
\eta_H=-\frac{\ddot{H}}{2\dot{H}H}\, .
\end{equation}
We can express the second Hubble slow-roll parameter $\eta_H$ as a function
of the field $\sigma $, by using the equations of motion (\ref{fieldhf}),
from which we obtain the following equations,
\begin{equation}\label{eqnsdfger}
\ddot{H}=-\kappa^2 \dot{\sigma} \ddot{\sigma }\, ,
\end{equation}
and by combining Eqs. (\ref{eqnsdfger}) and (\ref{voithitiki}), the Hubble
slow-roll parameter $\eta_H$ becomes,
\begin{equation}\label{sechubprolept}
\eta_H=-\frac{-3H\dot{\sigma}-\frac{\dot{\varphi}}{\dot{\sigma}}\tilde{V}_{,
\varphi}-\frac{\dot{\phi}}{\dot{\sigma}}\tilde{V}_{,\phi}}{\frac{\kappa^2}{3
}\dot{\sigma}\left(\dot{\sigma}^2+2\tilde{V}\right)}\, ,
\end{equation}
so for the potential of Eq. (\ref{vtilde}), the second Hubble slow-roll
parameter becomes,
\begin{align}\label{fdf}
& \eta_H=-\frac{3H }{\frac{\kappa^2}{3}\left(1+e^{-\sqrt{6} \kappa  \varphi
} \lambda (\phi )+2 \left(U(\varphi )-V(\phi ) e^{-2 \sqrt{\frac{3}{2}}
\kappa \varphi }+\lambda (\phi ) e^{-2 \sqrt{\frac{3}{2}} \kappa  \varphi
}\right)\right)}\\ \notag &
-\frac{\dot{\varphi}\left( \sqrt{6} e^{-\sqrt{6} \kappa  \varphi } \kappa
V(\phi )-\sqrt{6} e^{-\sqrt{6} \kappa  \varphi } \kappa  \lambda (\phi
)+U'(\varphi )\right)}{\left(1+e^{-\sqrt{6} \kappa  \varphi } \lambda (\phi
)\right)\frac{\kappa^2}{3}\left(1+e^{-\sqrt{6} \kappa  \varphi } \lambda
(\phi )+2 \left(U(\varphi )-V(\phi ) e^{-2 \sqrt{\frac{3}{2}} \kappa
\varphi }+\lambda (\phi ) e^{-2 \sqrt{\frac{3}{2}} \kappa  \varphi
}\right)\right)}\\ \notag &
-\frac{\dot{\phi}\left( -e^{-\sqrt{6} \kappa  \varphi } V'(\phi
)+e^{-\sqrt{6} \kappa  \varphi } \lambda '(\phi
)\right)}{\left(1+e^{-\sqrt{6} \kappa  \varphi } \lambda (\phi
)\right)\frac{\kappa^2}{3}\left(1+e^{-\sqrt{6} \kappa  \varphi } \lambda
(\phi )+2 \left(U(\varphi )-V(\phi ) e^{-2 \sqrt{\frac{3}{2}} \kappa \varphi
}+\lambda (\phi ) e^{-2 \sqrt{\frac{3}{2}} \kappa  \varphi
}\right)\right)}\, .
\end{align}
We need to note that the spectral index of primordial curvature
perturbations, in terms of the Hubble slow-roll parameters is equal to,
\begin{equation}\label{shedpectralperturb}
n_s=1-4\epsilon +2\eta_H \, ,
\end{equation}
and also note that at first order in the slow-roll expansion, the Hubble
slow-roll $\eta_H$ and the potential slow-roll parameter are related as
$n_H=-\epsilon+\eta$. For more details on this, see \cite{barrowslowroll}.
In addition, one can in principle use the following relation,
\begin{equation}\label{midnhoer}
\ddot{\sigma}=
\frac{\frac{\mathrm{d}G_{IJ}}{\mathrm{d}t}\dot{\phi}^I\dot{\phi}^J+G_{IJ}\ddot{\phi}^I\dot{\phi}^J+G_{IJ}\dot{\phi}^I\ddot{\phi}^J}{2\sqrt{G_{IJ}\dot{\phi}^I}\dot{\phi}^J}\, ,
\end{equation}
but we used the one appearing in Eq. (\ref{voithitiki}), which is equivalent
to some extent.

The general conclusion in this case too, is that the Jordan frame approach
for mimetic $F(R)$ gravity is more appealing when compared to the Einstein
frame approach, since in this case too, the Hubble slow-roll parameters are
affected by the Lagrange multiplier and the potential, and the resulting
theory has many similarities with the two scalar inflationary models. In
principle though, as we already stated in the previous section, one should
examine in detail the trajectory with $\phi\sim 0$, $\dot{\phi}\sim 0$,
which could be potentially interesting since it resembles the single scalar
field case, but it is a slightly deformed version.

\section{Dynamical System Analysis of Mimetic $F(R)$ gravity with Potential}

In this section we shall investigate in detail the behavior of the FRW
equations of mimetic $F(R)$ gravity, when these are seen as a dynamical
system. Particularly, we shall find the fixed points of the dynamical system
and we shall study the stability of the fixed points. Also, the physical
significance of the fixed points and also the possible connection of the
fixed points with some trajectory is also discussed in brief. For a recent
study on mimetic gravity dynamical systems, see \cite{mimetic2}.

We start off our analysis by recalling the FRW equations corresponding to
the mimetic $F(R)$ gravity which we quote here for convenience,
\begin{equation}\label{enm1slowdown}
-F(R)+6(\dot{H}+H^2)F'(R)-6H\frac{\mathrm{d}F'(R)}{\mathrm{d}t}-\lambda
(\dot{\phi}^2+1)+V(\phi)=0\, ,
\end{equation}
\begin{equation}\label{enm2slowdown}
F(R)-2(\dot{H}+3H^2)+2\frac{\mathrm{d}^2F'(R)}{\mathrm{d}t^2}+4H\frac{\mathrm{d}F'(R)}{\mathrm{d}t}-\lambda (\dot{\phi}^2-1)-V(\phi)=0\, ,
\end{equation}
\begin{equation}\label{enm3slowdown}
2\dot{\lambda}(t)+6H\lambda \dot{\phi}-\dot{V}(t)=0\, ,
\end{equation}
where in the last equation, namely Eq. (\ref{enm3slowdown}), we used the
fact that $\phi=t$. In the rest of this section we shall make extensive use
of the above equations. By introducing the following variables:
\begin{equation}\label{variablesslowdown}
x_1=-\frac{\dot{F'}(R)}{F'(R)H},\,\,\,x_2=-\frac{F(R)}{6F'(R)H^2},\,\,\,x_3=
\frac{R}{6H^2},\,\,\,x_4=\frac{V(t)}{6F'(R)H^2},\,\,\,x_5=-\frac{2\lambda(t)
}{6F'(R)H^2},\,\,\,x_6=\frac{1}{F'(R)}\, .
\end{equation}
Note that the variable $x_6$ takes values different from zero and it
measures in some way the deviation of the $F(R)$ gravity from the
$\Lambda$CDM gravity. Note that in Eqs. (\ref{enm1slowdown}),
(\ref{enm2slowdown}), (\ref{enm3slowdown}) and (\ref{variablesslowdown}),
the prime indicates differentiation with respect to the scalar curvature
$R$. By differentiating the variables (\ref{variablesslowdown}) with respect
to the $e$-folding number $N$, by making use of the rules of Eq.
(\ref{transfefold}), we obtain the following dynamical system,
\begin{align}\label{dynamicalsystemmain}
& \frac{\mathrm{d}x_1}{\mathrm{d}N}=-9x_4-x_3x_6-x_6-3x_2-x_1x_3-x_1^2\, ,
\\ \notag &
\frac{\mathrm{d}x_2}{\mathrm{d}N}=-m-4x_3+8+x_2x_1-2x_2x_3+4x_2 \, ,\\
\notag &
\frac{\mathrm{d}x_3}{\mathrm{d}N}=m+8x_3-2x_3^2-8 \, ,\\ \notag &
\frac{\mathrm{d}x_4}{\mathrm{d}N}=6x_5-22x_3-3x_1+4-6x_4-x_3x_6-x_6-3x_2+x_4
x_1-2x_4x_3+4x_4 \, ,\\ \notag &
\frac{\mathrm{d}x_5}{\mathrm{d}N}=-2x_3+4-3x_1-6x_4-x_3x_6-x_6-3x_2+x_1x_5-2
x_5x_3+4x_5 \, ,\\ \notag &
\frac{\mathrm{d}x_6}{\mathrm{d}N}=x_1x_6\, ,
\end{align}
where $m=\frac{\ddot{H}}{H^3}$. In the appendix $C$ we have added some
complementary equations that are useful for the derivation of the dynamical
system (\ref{dynamicalsystemmain}). The fixed points of the dynamical system
above are too many to quote here, but most of these can be considered as
unphysical or very restricted, since they lead to the solution
$x_6\rightarrow 0$, which has to hold true for all $N$ for a given $F(R)$
gravity. This condition restricts the $F(R)$ function, and therefore we
shall not analyze these points here.

The fixed points which can be easily analyzed are the following two, which
we denote as $P_1$ and $P_2$, with $P_1$ being:
\begin{align}\label{fixedpoints}
& P_1: x_1\to 0,\,\,\,\,\,x_2\to
-2+\frac{\sqrt{m}}{\sqrt{2}},\,\,\,\,\,x_3\to \frac{1}{2} \left(4-\sqrt{2}
\sqrt{m}\right),\,\,\,\,\,x_4\to \frac{-414 \sqrt{2} \sqrt{m}+60 m+79
\sqrt{2} m^{3/2}-22 m^2}{162-45 m+2 m^2},\\ \notag & x_5\to \frac{1080-90
\sqrt{2} \sqrt{m}-192 m+33 \sqrt{2} m^{3/2}-2 m^2}{162-45 m+2
m^2},\,\,\,\,\,x_6\to \frac{3 \left(108+405 \sqrt{2} \sqrt{m}+45 m-64
\sqrt{2} m^{3/2}+2 m^2\right)}{162-45 m+2 m^2}\, ,
\end{align}
while the point $P_2$ is,
\begin{align}\label{fixedpoints1}
& P_2: x_1\to 0,\,\,\,\,\,x_2\to \frac{1}{2} \left(-4-\sqrt{2}
\sqrt{m}\right),\,\,\,\,\,x_3\to \frac{1}{2} \left(4+\sqrt{2}
\sqrt{m}\right),\,\,\,\,\,x_4\to \frac{414 \sqrt{2} \sqrt{m}+60 m-79
\sqrt{2} m^{3/2}-22 m^2}{162-45 m+2 m^2},\\ \notag & x_5\to \frac{1080+90
\sqrt{2} \sqrt{m}-192 m-33 \sqrt{2} m^{3/2}-2 m^2}{162-45 m+2
m^2},\,\,\,\,\,x_6\to \frac{3 \left(108-405 \sqrt{2} \sqrt{m}+45 m+64
\sqrt{2} \, . m^{3/2}+2 m^2\right)}{162-45 m+2 m^2}
\end{align}
In addition, the effective equation of state (EoS) can be expressed in terms
of the variable $x_3$ and reads,
\begin{equation}\label{eos1}
w_{eff}=-\frac{1}{3} (2 x_3-1)\, ,
\end{equation}
so for the point $P_1$ it is equal to,
\begin{equation}\label{eos2}
w_{eff}=\frac{1}{3} \left(-3+\sqrt{2} \sqrt{m}\right)\, ,
\end{equation}
while for the point $P_2$ we have,
\begin{equation}\label{eos3}
w_{eff}=\frac{1}{3} \left(-3-\sqrt{2} \sqrt{m}\right) \, .
\end{equation}
Let us focus on the first fixed point $P_1$ and we shall investigate the
following three cases of interest, namely the cases that $w_{eff}$ describes
de Sitter expansion, radiation and cold dark-matter evolution. Before going
into this, we shall linearize the dynamic system by using the Hartman
linearization theorem, which we briefly discuss now. For an informative
account on the issues of dynamical systems which we shall use, see
\cite{jost}. The Hartman linearization theorem provides information with
regards to the stability and dynamical behavior of dynamical systems near
hyperbolic fixed points. Note that in the case of non-linear dynamical
systems, local stability of a fixed point does not ensure asymptotic
stability, an effect which is due to the non-linear part of the dynamical
system under study. Suppose that $\Phi (t)$ $\epsilon$ $R^n$ is the solution
to the dynamical system,
\begin{equation}\label{ds1}
\frac{\mathrm{d}\Phi}{\mathrm{d}t}=g(\Phi (t))\, ,
\end{equation}
where $g(\Phi (t))$ is locally Lipschitz continuous function
$g:R^n\rightarrow R^n$. Suppose that $\phi_*$ denotes the fixed points of
the dynamical system, for which the Jacobian matrix $\mathcal{J}(g)$ defined
to be,
\begin{equation}\label{jaconiab}
\mathcal{J}=\sum_i\sum_j\Big{[}\frac{\mathrm{\partial f_i}}{\partial
x_j}\Big{]}\,
\end{equation}
at these points has eigenvalues $e_i$ that satisfy $\mathrm{Re}(e_i)\neq 0$.
We shall denote the spectrum of the eigenvalues of a matrix $A$, as $\sigma
(A)$, so a global hyperbolic fixed point satisfies $\mathrm{Re}\left(\sigma
(\mathcal(J))\right)\neq 0$. Hartman's theorem states that there exists a
homeomorphism $\mathcal{F}:U\rightarrow R^n$, where $U$ is a small open
neighborhood of a fixed point $\phi_*$, which is defined in such a way so
that $\mathcal{F}(\phi_*)$. This homeomorphism has a flow
$\frac{\mathrm{d}h(u)}{\mathrm{d}t}$, which is equal to,
\begin{equation}\label{fklow}
\frac{\mathrm{d}h(u)}{\mathrm{d}t}=\mathcal{J}h(u)\, ,
\end{equation}
and also it is topologically conjugate to the flow of Eq. (\ref{ds1}). This
theorem holds true even for non-autonomous dynamical systems, like the ones
we shall deal in the present section. In simple words, this means that the
dynamical system of Eq. (\ref{ds1}) takes the following form,
\begin{equation}\label{dapprox}
\frac{\mathrm{d}\Phi}{\mathrm{d}t}=\mathcal{J}(g)(\Phi)\Big{|}_{\Phi=\phi_*}
(\Phi-\phi_*)+\mathcal{S}(\phi_*,t)\, ,
\end{equation}
with $\mathcal{S}(\phi,t)$ a smooth map $[0,\infty )\times R^n$. As a
consequence of the theorem, if for the matrix $\mathcal{J}(g)$ holds true
that $\mathcal{Re}\left(\sigma (\mathcal{J}(g))\right)<0$, and also if ,
\begin{equation}\label{gfgd}
\lim_{\Phi\rightarrow
\phi_*}\frac{|\mathcal{S}(\phi,t)|}{|\Phi-\phi_*|}\rightarrow 0\, ,
\end{equation}
then, the fixed point $\phi_*$ of the dynamical flow
$\frac{\mathrm{d}\Phi}{\mathrm{d}t}=\mathcal{J}(g)(\Phi)\Big{|}_{\Phi=\phi_*
}(\Phi-\phi_*)$ is an asymptotically stable fixed point of the flow given in
Eq. (\ref{dapprox}). If any of the above conditions are violated, then extra
conditions are needed to decide whether the fixed point is stable or
unstable. Some insights on deciding whether de Sitter points for certain
$F(R)$ functions are stable or not were provided in an earlier section in
this paper. In this section we shall be interested in the fixed points of
the FRW equations, seen as a dynamical system, and we also critically
discuss the stability of these fixed points using the Hartman theorem.

In the case of the dynamical system of Eq. (\ref{dynamicalsystemmain}), the
matrix $\mathcal{J}=\sum_i\sum_j\Big{[}\frac{\mathrm{\partial f_i}}{\partial
x_j}\Big{]}$ is,
\begin{equation}\label{matrixceas}
\mathcal{J}=\left(
\begin{array}{cccccc}
 -2 x_1-x_3 & -3 & -x_1-x_6 & -9 & 0 & -1-x_3
\\
 x_2 & 4+x_1-2 x_3 & -4-2 x_2 & 0 & 0 & 0 \\
 0 & 0 & 8-4 x_3 & 0 & 0 & 0 \\
 -3+x_4 & -3 & -22-2 x_4-x_6 & -2+x_1-2 x_3 &
6 & -1-x_3 \\
 -3+x_5 & -3 & -2-2 x_5-x_6 & -6 & 4+x_1-2 x_3
& -1-x_3 \\
 x_6 & 0 & 0 & 0 & 0 & x_1 \\
\end{array}
\right)\, ,
\end{equation}
where obviously the $f_i$'s are,
\begin{align}\label{fis}
& f_1=-9x_4-x_3x_6-x_6-3x_2-x_1x_3-x_1^2\, , \\ \notag &
f_2=-m-4x_3+8+x_2x_1-2x_2x_3+4x_2 ,\\ \notag &
f_3=m+8x_3-2x_3^2-8 \, ,\\ \notag &
f_4=6x_5-22x_3-3x_1+4-6x_4-x_3x_6-x_6-3x_2+x_4x_1-2x_4x_3+4x_4 \, ,\\ \notag
&
f_5=-2x_3+4-3x_1-6x_4-x_3x_6-x_6-3x_2+x_1x_5-2x_5x_3+4x_5\, , \\ \notag &
f_6=x_1x_6 \, .
\end{align}
By using the values for the parameters $x_i$, $=1,...6$ given in Eq.
(\ref{fixedpoints}), and by substituting in Eq. (\ref{matrixceas}) we may
easily obtain the eigenvalues $j_i$ of $\mathcal{J}$, which are the
following,
\begin{equation}\label{eigenv1}
j_0=8-2 \left(4-\sqrt{2} \sqrt{m}\right),\,\,\, z_i,\,\,i=1,...5\, ,
\end{equation}
where the $z_i$ are the solutions to the following algebraic equation,
\begin{align}\label{alfgereeqns}
& 216 \sqrt{2} \sqrt{m}+1476 m-402 \sqrt{2} m^{3/2}+12 m^2-1332 \sqrt{2}
\sqrt{m} z+834 m z-32 \sqrt{2} m^{3/2} z+4 m^2 z+540 z^2\\ \notag & -258
\sqrt{2} \sqrt{m} z^2+78 m z^2-10 \sqrt{2} m^{3/2} z^2+42 z^3-42 \sqrt{2}
\sqrt{m} z^3+18 m z^3+16 z^4-7 \sqrt{2} \sqrt{m} z^4+2 z^5=0\, ,
\end{align}
which can be found numerically, when $m$ is specified. So let us investigate
which values of $m$ give the de Sitter, the matter and the radiation era.
Assume for simplicity that $m$ is a constant parameter. The de Sitter era
corresponds to $w_{eff}=-1$, so by using Eq. (\ref{eos2}), we can see that
the de Sitter era corresponds to $m=0$, the matter dominated era to $m=9/2$,
while the radiation era corresponds to $m=8$. For the de Sitter era, the
fixed point $P_1$ becomes,
\begin{align}\label{fixedpointsdessitter}
& P_1: x_1\to 0,\,\,\,\,\,x_2\to -2,\,\,\,\,\,x_3\to 2 ,\,\,\,\,\,x_4\to
0,\,\,\, x_5\to 20/3,\,\,\,\,\,x_6\to 2\, ,
\end{align}
and the eigenvalues of the matrix $\mathcal{J}$ are equal to,
\begin{equation}\label{dgfgd}
\left\{0,-9,0,0,\frac{1}{2} \left(1-i \sqrt{119}\right),\frac{1}{2}
\left(1+i \sqrt{119}\right)\right\}\, .
\end{equation}
Hence the matrix $\mathcal{J}$ has some eigenvalues zero and also some other
have positive real part. We could say that this indicates instability, but
one cannot be sure, since Hartman's theorem does not apply. Another reason
against the applicability of Hartman's theorem is that the ``perturbation
matrix'', which we denoted as $\mathcal{S}$ in Eq. (\ref{dapprox}), does not
satisfy the condition (\ref{gfgd}), when the dynamical system
(\ref{dynamicalsystemmain}) is considered. Indeed, for the evolution
(\ref{dynamicalsystemmain}), the matrix $\mathcal{S}$ is equal to,
\begin{equation}\label{msmathcals}
\mathcal{S}=
\left(%
\begin{array}{c}
  -x_3 x_6-x_1 x_3-x_1^2 \\
  -m+8+x_2 x_1-2 x_2 x_3 \\
  m -2x_3^2-8 \\
  4-x_3 x_6+x_4 x_1-2 x_4 x_3 \\
  4-x_3 x_6+x_1 x_5-2x_5 x_3 \\
  x_1 x_6 \\
\end{array}%
\right)
\end{equation}
and it easy to see that for $m=0$, the condition (\ref{gfgd}) is not
satisfied. Hence further information is needed in order to decide if this de
Sitter point is an attractor (stability or asymptotic stability of the fixed
point), or unstable. One should for example search for the center manifolds
and sub-manifolds \cite{jost}, or address the problem by using Lyapunov's
theorems. This task however is beyond the scopes of this paper, and we hope
to address in the future. We now proceed to the radiation fixed point for
which $w_{eff}=\frac{1}{3}$, so Eq. (\ref{eos2}) gives $m=8$, for which
value, the fixed point $P_1$ becomes,
\begin{align}\label{fixedpointsradiation}
& P_1: x_1\to 0,\,\,\,\,\,x_2\to 0,\,\,\,\,\,x_3\to 0 ,\,\,\,\,\,x_4\to
4/5,\,\,\, x_5\to -8/5,\,\,\,\,\,x_6\to -36/5\, ,
\end{align}
and the corresponding eigenvalues of the matrix $\mathcal{J}$ are equal to,
\begin{equation}\label{dgfgd}
\{8,-0.778,4,5.382,-1.301-3.934\text{  }i,-1.301+3.934\text{} i\}\, .
\end{equation}
By substituting the eigenvalues to the matrix $\mathcal{S}$ we can see that
the matrix $\mathcal{S}$ does not satisfy Hartman's conditions, so for this
case too, further information and analysis is needed. The same applies for
the matter domination point, for which $m=9/2$, but we omit the analysis for
brevity, since the resulting picture is qualitatively the same. The only
notable feature of the matter fixed point ($w_{eff}=0$), is that for
$m=9/2$, the fixed point $P_1$ becomes,
\begin{equation}\label{sheltersaskia}
 P_1: x_1\to 0,\,\,\,\,\,x_2\to -\frac{1}{2},\,\,\,\,\,x_3\to \frac{1}{2}
,\,\,\,\,\,x_4\to \infty,\,\,\, x_5\to \infty,\,\,\,\,\,x_6\to \infty\, ,
\end{equation}
so interestingly enough the variables $x_4$ and $x_5$, which are related to
the mimetic potential and Lagrange multiplier, take infinite values near the
$w_{eff}=0$ evolution, always for constant $m$. Before ending the study for
the fixed point $P_1$, we need to note that no constant $m$ exists, for
which the phantom evolution can be described by the fixed point $P_1$. As we
now demonstrate, the fixed point $P_2$ describes the phantom evolution, for
constant $m$.

We now turn our focus on the second fixed point $P_2$ appearing in Eq.
(\ref{fixedpoints1}), and we study the behavior for various values of the
parameter $m$ and address the stability of the fixed point. Intriguingly
enough, when $m$ is constant, the second fixed point cannot describe any
ordinary matter fluid evolution, like for example cold dark matter
($w_{eff}=0$) or radiation ($w_{eff}=1/3$), but can describe de Sitter
evolution for $m=0$ and phantom evolution, for $m>0$. Let us investigate the
de Sitter point, which occurs for $m=0$. The fixed point $P_2$ for $m=0$
becomes
\begin{align}\label{fixeddsfp2}
& P_2: x_1\to 0,\,\,\,\,\,x_2\to -2,\,\,\,\,\,x_3\to 2 ,\,\,\,\,\,x_4\to
0,\,\,\, x_5\to \frac{20}{3},\,\,\,\,\,x_6\to 2\, ,
\end{align}
and the eigenvalues of the matrix $\mathcal{J}$ for the fixed point $P_2$ at
$m=0$ are equal to,
\begin{equation}\label{dgfgd}
\left\{0,-9,0,0,\frac{1}{2} \left(1-i \sqrt{119}\right),\frac{1}{2}
\left(1+i \sqrt{119}\right)\right\}      \, .
\end{equation}
which are identical to the ones corresponding to the de Sitter point of the
fixed point $P_1$. The same stability criteria apply in this case too, so we
proceed to the case that $m>0$. Note that the case $m<0$ yields complex
values for the fixed points, so we omit this study. For the phantom
evolution, consider for example that $w_{eff}=-1.1$, which corresponds to
$m=0.045$. In this case, the fixed point $P_2$ becomes,
\begin{align}\label{fixeddsfp2phantom}
& P_2: x_1\to 0,\,\,\,\,\,x_2\to -2.15,\,\,\,\,\,x_3\to 2.15
,\,\,\,\,\,x_4\to 0.786284,\,\,\, x_5\to 6.86284,\,\,\,\,\,x_6\to
-0.198906\, ,
\end{align}
and the eigenvalues of the matrix $\mathcal{J}$ are,
\begin{equation}\label{eigenvmatrj}
\{-0.6,-9.36598,-0.725613,-0.0165485,0.529073\, -5.58193 i,0.529073\,
+5.58193 i\}\, .
\end{equation}
By calculating the matrix $\mathcal{S}$ and taking the limit $x_i\rightarrow
0$, we can see that the Hartman criteria are not satisfied, so the linear
stability theorems do not suffice for deciding whether the fixed points are
stable or unstable.

In conclusion, the study revealed two quite interesting fixed points with
physical significance, since if the fixed point $P_1$ describes an unstable
fixed point corresponding to some era and the point $P_2$ for the same value
of $m$ is a stable point (since $m>0$), then this point might be the
attractor of every solution of the dynamical system, so in the end, the
points $P_1$ and $P_2$ might be connected. For example, the matter era
described by the point $P_1$ may end up to the phantom evolution of the
fixed point $P_2$. Also we might connect the two de Sitter points
corresponding to the points $P_1$ and $P_2$. Therefore, the stability issues
of the fixed points should be further addressed, by using more involved
methods compared to the linearization theorems. We hope to address in detail
this issue in the future.

Before we close this section, it is worth giving a brief account of the case
that perfect matter fluids are also taken into account. Consider for example
the case that the perfect fluid added has an equation of state
$P_w=w\rho_w$, with $w$ its equation of state and $P_w$, $\rho_w$ are its
effective pressure and effective density. Introducing the new variable $x_7$
in the equations of motion, the dynamical system of Eq.
(\ref{dynamicalsystemmain}) becomes,
\begin{align}\label{dynamicalsystemmainwithmatter}
&
\frac{\mathrm{d}x_1}{\mathrm{d}N}=-9x_4-x_3x_6-x_6-3x_2-x_1x_3-x_1^2+3\,w\,x
_7\, , \\ \notag &
\frac{\mathrm{d}x_2}{\mathrm{d}N}=-m-4x_3+8+x_2x_1-2x_2x_3+4x_2 \, ,\\
\notag &
\frac{\mathrm{d}x_3}{\mathrm{d}N}=m+8x_3-2x_3^2-8 \, ,\\ \notag &
\frac{\mathrm{d}x_4}{\mathrm{d}N}=6x_5-22x_3-3x_1+4-6x_4-x_3x_6-x_6-3x_2+x_4
x_1-2x_4x_3+4x_4-3 (1+w)x_7 \, ,\\ \notag &
\frac{\mathrm{d}x_5}{\mathrm{d}N}=-2x_3+4-3x_1-6x_4-x_3x_6-x_6-3x_2+x_1x_5-2
x_5x_3+4x_5-3(1+w)w\,x_7 \, ,\\ \notag &
\frac{\mathrm{d}x_6}{\mathrm{d}N}=x_1x_6\, ,\\ \notag &
\frac{\mathrm{d}x_7}{\mathrm{d}N}=-3\,(1+w)\, x_7+x_7x_1-2x_7x_3+4x_7\, .
\end{align}
We can easily obtain the fixed points of the dynamical system
(\ref{dynamicalsystemmainwithmatter}), which can be found in the appendix
$D$. As one can see, the points $P_1$ and $P_2$ are identical to the fixed
points $P_1$ and $P_2$ of the perfect fluid-free dynamical system of Eq.
(\ref{dynamicalsystemmain}), with the addition that the extra variable
corresponds to $x_7\rightarrow 0$. In addition to these two, there are many
fixed points that lead to the condition $x_7\rightarrow 0$ and
$x_6\rightarrow 0$, which are also identical to the corresponding ones of
the dynamical system (\ref{dynamicalsystemmain}). The new fixed points $P_3$
and $P_4$ (see appendix D), lead to the condition $x_6\rightarrow 0$, which
severely restricts the $F(R)$ gravity, since it has to hold true for all
$N$. We therefore defer this study to a future publication and we focus on
the points $P_1$ and $P_2$. As we now show, the eigenvalues of the matrix
$\mathcal{J}$ change and therefore it is expected that the stability
conditions will change. However, the linearization theorems do not apply in
this case, so the matter fluid case must be addressed by using more involved
techniques for dynamical systems. Before closing, we shall give the
resulting form of the matrix $\mathcal{J}$, which is,
\begin{equation}\label{mathcaljmatrixmatter}
\mathcal{J}=\left(
\begin{array}{ccccccc}
 -2 x_1-x_3 & -3 & -x_1-x_6 & -9 & 0 & -1-x_3
& 3 w \\
 x_2 & 4+x_1-2 x_3 & -4-2 x_2 & 0 & 0 & 0 & 0 \\
 0 & 0 & 8-4 x_3 & 0 & 0 & 0 & 0 \\
 -3+x_4 & -3 & -22-2 x_4-x_6 & -2+x_1-2 x_3 &
6 & -1-x_3 & -3 (1+w) \\
 -3+x_5 & -3 & -2-2 x_5-x_6 & -6 & 4+x_1-2 x_3
& -1-x_3 & -3 w (1+w) \\
 x_6 & 0 & 0 & 0 & 0 & x_1 & 0 \\
 x_7 & 0 & -2 x_7 & 0 & 0 & 0 & 4-3 (1+w)+x_1-2 x_3
\\
\end{array}
\right)\, .
\end{equation}
Finally, the eigenvalues of $\mathcal{J}$ for the point $P_1$ are equal to,
\begin{equation}\label{eigemattr1}
\{8-2 \left(4-\sqrt{2} \sqrt{m}\right),\sqrt{2} \sqrt{m}-3 (1+w),z_i\}
\end{equation}
with $z_i$ the roots of the algebraic equation,
\begin{align}\label{frikoula}
& 216 \sqrt{2} \sqrt{m}+1476 m-402 \sqrt{2} m^{3/2}+12 m^2-1332 \sqrt{2}
\sqrt{m} z+834 m z-32 \sqrt{2} m^{3/2} z+4 m^2 z+540 z^2\\ \notag & -258
\sqrt{2} \sqrt{m} z^2+78 m z^2-10 \sqrt{2} m^{3/2} z^2+42 z^3-42 \sqrt{2}
\sqrt{m} z^3+18 m z^3+16 z^4-7 \sqrt{2} \sqrt{m} z^4+2 z^5=0\, .
\end{align}
For the point $P_2$, the eigenvalues are,
\begin{equation}\label{eigemattr1sdsd}
\{8-2 \left(4+\sqrt{2} \sqrt{m}\right),-\sqrt{2} \sqrt{m}-3 (1+w),z_i\}
\end{equation}
with $z_i$ the roots of the algebraic equation,
\begin{align}\label{frikoulasdsdsds}
& -216 \sqrt{2} \sqrt{m}+1476 m+402 \sqrt{2} m^{3/2}+12 m^2+1332 \sqrt{2}
\sqrt{m} z+834 m z+32 \sqrt{2} m^{3/2} z+4 m^2 z\\ \notag & +540 z^2+258
\sqrt{2} \sqrt{m} z^2+78 m z^2+10 \sqrt{2} m^{3/2} z^2+42 z^3+42 \sqrt{2}
\sqrt{m} z^3+18 m z^3+16 z^4+7 \sqrt{2} \sqrt{m} z^4+2 z^5=0\, .
\end{align}
We shall not compute the corresponding eigenvalues for the fixed points
$P_3$ and $P_4$, since this case is more complicated and leads to a very
restricted form of the $F(R)$ function. This analysis will be performed to a
much more focused paper, so we defer this to a future publication.

\section{Inverse Reconstruction Method for the $F(R)$ Gravity}

In this section we describe another reconstruction method which by providing
the Hubble rate, the potential $V(\phi )$ and the Lagrange multiplier
$\lambda (\phi )$, it enables us to find the $F(R)$ gravity which generates
such a cosmological evolution. We exemplify this by using two of the
bouncing cosmologies we studied earlier in this paper, the matter bounce
scenario of Eq. (\ref{mattbouncescenarohib}) and the superbounce scenario of
Eq. (\ref{superbouncescenariob}). We start of with the matter bounce
scenario, and by using Eq. (\ref{diffeqnnewaddition}) it is possible to find
which $F(R)$ gravity can generate the superbounce cosmology, for a suitably
chosen mimetic potential $V(t)$ and Lagrange multiplier $\lambda (t)$. In
principle, these can be arbitrarily chosen, but a non-convenient choice may
render the differential equation (\ref{diffeqnnewaddition}) quite difficult
to solve analytically.

We start by studying the matter bounce scenario, with Hubble rate given in
(\ref{mattbouncescenarohib}), and a convenient choice for the mimetic
potential $V(\phi=t)$ and the Lagrange multiplier $\lambda (t)$ is the
following,
\begin{equation}\label{potandlag}
V(t)=-\frac{32 t^2 \mu ^2}{3 \left(1+t^2 \mu \right)^2},\, \, \, \lambda
(t)=\frac{8 t^2 \mu ^2}{3 \left(1+t^2 \mu \right)^2}-\frac{4 \mu }{3
\left(1+t^2 \mu \right)}\, .
\end{equation}

Having at hand the cosmological evolution (\ref{mattbouncescenarohib}), the
mimetic potential and the Lagrange multiplier given in Eq.
(\ref{potandlag}), we may easily find the $F(R)$ gravity that realizes such
a cosmological scenario. Indeed, by combining Eqs.
(\ref{diffeqnnewaddition}), (\ref{scalarpot}) and (\ref{lagrange}), we
acquire the following differential equation,
\begin{equation}\label{diifequation}
8 \mu  \left(3+t^2 \mu \right) y(t)+2 \left(1+t^2 \mu \right) \left(-4 t \mu
y'(t)+3 \left(1+t^2 \mu \right) y''(t)\right)=0\, ,
\end{equation}
with $y(t)=F'(R(t))$, and by solving this with respect to $y(t)$, we obtain
the following solution,
\begin{equation}\label{finalfrsolution}
y(t)=\frac{\left(1+i t \sqrt{\mu }\right)^{7/6} \left(3-5 t^2 \mu \right)
\left(1+t^2 \mu \right)^{5/6} C_1}{8 \left(1+\frac{1}{2} \left(-1+i t
\sqrt{\mu }\right)\right)^{7/3} \left(1-i t \sqrt{\mu }\right)^{7/6}
\Gamma\left(-\frac{4}{3}\right)}+\left(1+t^2 \mu \right)^{5/6} C_2
Q_{n_1}^{m_1}\left(i t \sqrt{\mu }\right)\, ,
\end{equation}
where $n_1=-\frac{1}{3}$, $m_1=\frac{7}{3}$ and $C_1$, $C_2$ arbitrary
constants.  Also $Q_{n_1}^{m_1}$ is the associated Legendre function of the
second kind. Bearing in mind that $R=6 \dot{H}+12 H^2$, substituting to this
equation the Hubble rate (\ref{mattbouncescenarohib}), and by solving with
respect to $t$, we get,
\begin{equation}\label{dgfhd}
t=\sqrt{\frac{20}{3 R}-\frac{1}{\mu }+\frac{4 \sqrt{-3 R \mu ^3+25 \mu
^4}}{3 R \mu ^2}} \, .
\end{equation}
Substituting this in Eq. (\ref{finalfrsolution}), we obtain the function
$y(R)$ and then we can have the analytic form of $F'(R)$ in the large and
small curvature limits. For example, in the large curvature limit, the
$F'(R)$ gravity is approximately equal to,
\begin{align}\label{approxfr}
& F'(R)\simeq \frac{2 i 2^{2/3} \pi ^2 \left(-\frac{1}{\mu }\right)^{3/2}
\mu ^{3/2} \left(\frac{\sqrt{-\mu ^3}}{\mu }\right)^{2/3} C_2 \left(2
\Gamma\left(-\frac{4}{3}\right)-\Gamma\left(-\frac{1}{3}\right)\right)
\left(\frac{1}{R}\right)^{1/3}}{3\ 3^{5/6} \Gamma\left(-\frac{4}{3}\right)
\Gamma\left(-\frac{5}{6}\right) \Gamma\left(-\frac{1}{3}\right)
\Gamma\left(\frac{11}{6}\right)}\\ \notag & +
\frac{2 i 2^{2/3} \pi ^2 \left(-\frac{1}{\mu }\right)^{3/2} \mu ^{3/2} C_2
R^{1/6}}{3^{1/3} \left(\frac{\sqrt{-\mu ^3}}{\mu }\right)^{1/3}
\Gamma\left(-\frac{4}{3}\right) \Gamma\left(-\frac{5}{6}\right)
\Gamma\left(\frac{11}{6}\right)}\, ,
\end{align}
We can see that the $F(R)$ gravity is complex, and let us briefly recall
what this indicated in pure $F(R)$ gravity theory. As we will see, in
mimetic $F(R)$ gravity, the physical significance of a complex $F(R)$
gravity can be entirely different in comparison to pure $F(R)$ gravity.
Indeed, as was demonstrated in Ref. \cite{sergeicomplex}, phantom scalar
models and complex $F(R)$ gravity are mathematically equivalent, at least,
in rip point. In addition, even in ordinary non-phantom scalar-tensor
theory, in regions were the scalar-tensor theory develops a Big Rip
singularity \cite{bigrip}, the corresponding $F(R)$ gravity may become
complex. In principle, the complex $F(R)$ gravity cosmology can develop
quite intriguing features, with the most sound one being the inability to
analytically continue the Lorentz signature metric to a metric with a
Euclidean signature, in the Jordan frame. Also, as shown in
\cite{sergeioikonomou1,sergeioikonomou2,sergeioikonomou3,sergeioikonomou4},
bouncing cosmologies in the Jordan frame are never generated by complex
$F(R)$ gravity and also bouncing cosmologies in the Einstein frame never
lead to complex $F(R)$ gravity in the Jordan frame. However, as we
demonstrated in this section, within the mimetic $F(R)$ framework, a complex
$F(R)$ gravity in the Jordan frame may realize the matter bounce scenario,
for suitably chosen mimetic potential and Lagrange multiplier. We believe
that this feature of mimetic $F(R)$ gravity in the Jordan frame is quite
appealing, since it provides a completely different viewpoint of $F(R)$
gravity.

Before closing this section, we shall investigate from which $F(R)$ gravity
can the superbounce cosmological evolution of Eq.
(\ref{superbouncescenariob}) be realized in the context of mimetic $F(R)$
gravity. We choose the mimetic potential $V(t)$ and the Lagrange multiplier
$\lambda (t)$ to be equal to,
\begin{equation}\label{potandlaggfg}
V(t)=-\frac{24}{c^4 (t-t_s)^2},\, \, \, \lambda (t)=\frac{2}{c^2
(t-t_s)^2}\, .
\end{equation}
Following the same steps as in the previous example, we obtain the following
differential equation,
\begin{equation}\label{diifequationgfg}
2 \left(-6 \left(-2+c^2\right) y(t)+c^2 (t-t_s) \left(-2 y'(t)+c^2 (t-t_s)
y''(t)\right)\right)=0\, ,
\end{equation}
where again $y(t)=F'(R(t))$. By solving this with respect to $y(t)$, we
obtain the following solution,
\begin{equation}\label{finalfrsolutiongfg}
y(t)=C_1(t-t_s)^{\gamma_1}+C_2(t-t_s)^{\gamma_2}\, ,
\end{equation}
where $C_1$, $C_2$ are again some arbitrary constants, and the parameters
$\gamma_1,\gamma_2$ appear in the Appendix A. By substituting in $R=6
\dot{H}+12 H^2$, the Hubble rate (\ref{superbouncescenariob}) and solving
the resulting algebraic equation with respect to $t$, we get,
\begin{equation}\label{dgfhdgfg}
t=\frac{2 \sqrt{3} \sqrt{4 c^4 R-c^6 R}+c^4 R t_s}{c^4 R}\, .
\end{equation}
Substituting this in Eq. (\ref{finalfrsolution}), we can obtain the analytic
form of $y(R)$, and by integrating once with respect to $R$, we obtain the
$F(R)$ gravity. By doing so, the final form of the $F(R)$ gravity is equal
to,
\begin{align}\label{approxfrgfg}
& F(R)= \frac{C_1 R \left(-t_s+\frac{2 \sqrt{3} \sqrt{4 c^4 R-c^6 R}+c^4 R
t_s}{c^4 R}\right)^{\gamma_1}}{1-\frac{\gamma_1}{2}}+\frac{C_2 R
\left(-t_s+\frac{2 \sqrt{3} \sqrt{4 c^4 R-c^6 R}+c^4 R t_s}{c^4
R}\right)^{\gamma_2}}{1-\frac{\gamma_2}{2}}\, .
\end{align}
Hence, using the theoretical framework of mimetic $F(R)$ gravity, we
presented a reconstruction method for finding which $F(R)$ gravity can
generate a specific cosmological evolution, given the mimetic potential and
the Lagrange multiplier. In principle the mimetic potential and Lagrange
multiplier can arbitrarily be chosen, and hence, for a given cosmological
evolution, there are many different sets of $F(R)$, $V(\phi)$ and $\lambda
(\phi)$, constrained by Eq. (\ref{diffeqnnewaddition}), that can realize
this cosmological evolution.

Before closing this section, we shall present a particularly interesting
example to support the usefulness of the method we presented. Our aim is to
find a particular cosmology with Hubble rate $H(t)$ which ends up to the
$R^2$ model of inflation. So what we aim is to find the potential $V(t)$ and
the Lagrange multiplier $\lambda (t)$, which lead to $y(R)=1+\alpha R$, and
therefore the resulting $F(R)$ gravity is $F(R)=R+\alpha_1 R^2$.

\begin{figure}[h]
\centering
\includegraphics[width=16pc]{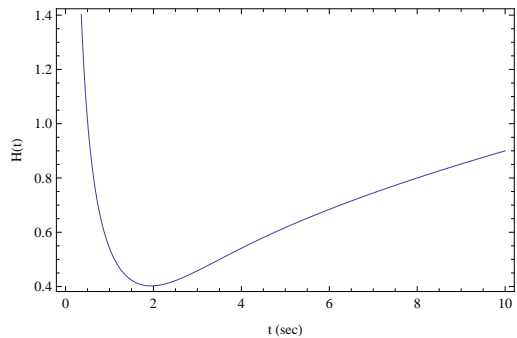}
\caption{Plot of the Hubble rate
$H(t)=\frac{\sqrt{t}I_{-\frac{2}{3}}(\frac{2 t^{3/2}}{3
\sqrt{3}})}{2\sqrt{3}I_{\frac{1}{3}}(\frac{2 t^{3/2}}{3 \sqrt{3}})}$.}
\label{hubfig}
\end{figure}

Consider for example that the Hubble rate is of the following form,
\begin{equation}\label{bessel}
 H(t)=\frac{\sqrt{t}I_{-\frac{2}{3}}(\frac{2 t^{3/2}}{3
\sqrt{3}})}{2\sqrt{3}I_{\frac{1}{3}}(\frac{2 t^{3/2}}{3 \sqrt{3}})}\, ,
\end{equation}
with $I_n(x)$ the modified Bessel function of first kind. As it can be seen
in Fig. \ref{hubfig} and can be easily crosschecked, the Hubble rate is
positive for all cosmic time values $t$, and no singularity issues occur.
By making the following choice for the mimetic scalar potential $V(t)$
(recall that we identified $\phi=t$),
\begin{align}\label{potestarobinve}
& V(t)=-\frac{t I_{-\frac{2}{3}}(\frac{2 t^{3/2}}{3 \sqrt{3}})^2}{2
I_{\frac{1}{3}}(\frac{2 t^{3/2}}{3 \sqrt{3}})^2}+6 \left(1+t \alpha \right )
\Big{(}\frac{t I_{-\frac{2}{3}}(\frac{2 t^{3/2}}{3 \sqrt{3}})^2}{12
I_{\frac{1}{3}}(\frac{2 t^{3/2}}{3
\sqrt{3}})^2}+\frac{I_{-\frac{2}{3}}(\frac{2 t^{3/2}}{3 \sqrt{3}})}{4
\sqrt{3} \sqrt{t} I_{\frac{1}{3}}(\frac{2 t^{3/2}}{3 \sqrt{3}})}+\frac{t
\left(I_{-\frac{5}{3}}(\frac{2 t^{3/2}}{3 \sqrt{3}})+I_{\frac{1}{3}}(\frac{2
t^{3/2}}{3 \sqrt{3}})\right)}{12 I_{\frac{1}{3}}(\frac{2 t^{3/2}}{3
\sqrt{3}})}\\ \notag &
-\frac{t I_{-\frac{2}{3}}(\frac{2 t^{3/2}}{3 \sqrt{3}})
\left(I_{-\frac{2}{3}}(\frac{2 t^{3/2}}{3 \sqrt{3}})+I_{\frac{4}{3}}(\frac{2
t^{3/2}}{3 \sqrt{3}})\right)}{12 I_{\frac{1}{3}}(\frac{2 t^{3/2}}{3
\sqrt{3}})^2}
\Big{)}
\end{align}
and also by choosing the Lagrange multiplier as follows,
\begin{align}\label{lagstarobla}
& \lambda (t)=-\frac{I_{-\frac{2}{3}}(\frac{2 t^{3/2}}{3 \sqrt{3}})}{4
\sqrt{3} \sqrt{t} I_{\frac{1}{3}}(\frac{2 t^{3/2}}{3
\sqrt{3}})}+\frac{\sqrt{t} \alpha I_{-\frac{2}{3}}(\frac{2 t^{3/2}}{3
\sqrt{3}})}{\sqrt{3} I_{\frac{1}{3}}(\frac{2 t^{3/2}}{3
\sqrt{3}})}-\frac{\sqrt{3} \sqrt{t} \alpha I_{-\frac{2}{3}}(\frac{2
t^{3/2}}{3 \sqrt{3}})}{2 I_{\frac{1}{3}}(\frac{2 t^{3/2}}{3 \sqrt{3}})}
\frac{t \left(I_{-\frac{5}{3}}(\frac{2 t^{3/2}}{3
\sqrt{3}})+I_{\frac{1}{3}}(\frac{2 t^{3/2}}{3 \sqrt{3}})\right)}{12
I_{\frac{1}{3}}(\frac{2 t^{3/2}}{3 \sqrt{3}})}\\ \notag &+\frac{t
I_{-\frac{2}{3}}(\frac{2 t^{3/2}}{3 \sqrt{3}})
\left(I_{-\frac{2}{3}}(\frac{2 t^{3/2}}{3 \sqrt{3}})+I_{\frac{4}{3}}(\frac{2
t^{3/2}}{3 \sqrt{3}})\right)}{12 I_{\frac{1}{3}}(\frac{2 t^{3/2}}{3
\sqrt{3}})^2}
\end{align}
we obtain the following analytic solution,
\begin{equation}\label{analyticfrdot}
y(R)=1+\alpha R\,
\end{equation}
where we made use of the fact that,
\begin{equation}\label{trsolution}
R=6\left (\dot{H}+2H^2\right )=t\, ,
\end{equation}
which easily follows by using the modified Bessel function properties and
the Hubble rate (\ref{bessel}). Therefore, the resulting $F(R)$ gravity
is $F(R)=R+\frac{\alpha}{2}R^2$, which shows us that the mimetic $F(R)$
framework in the Jordan frame results to a completely different picture in
comparison to the standard Jordan frame $F(R)$ gravity. As we just
demonstrated, the Jordan frame $R^2$ gravity with the potential and Lagrange
multiplier functions chosen as in (\ref{potestarobinve}) and
(\ref{lagstarobla}), realize the cosmology with Hubble rate (\ref{bessel}).

\section{$R^{1+\epsilon}$ Mimetic Gravity Cosmology and the Simplicity Argument for Physical Theories}

In the previous sections we demonstrated that in the context of mimetic $F(R)$ modified gravity with Lagrange multiplier and scalar potential, it is possible to generate various cosmological evolutions which are compatible with the recent Planck observational data. As we explicitly showed, any $F(R)$ gravity can yield results compatible with observations, by choosing the potential and the Lagrange multiplier appropriately. We need to note that the resulting forms for the potential and the Lagrange multiplier were quite complicated. 

Then an important fundamental question rises, since every $F(R)$ gravity is allowed by appropriately choosing the potential and the Lagrange multiplier, how should the correct model be singled out from the rest of the models? In addition, another question closely related to the first one is, what is the reason to have so complicated potential and Lagrange multiplier? Should the correct cosmological description be so complicated? 

These questions are of fundamental importance and we discuss these in detail in this section. To start with, let us answer the questions from a quantitative point of view firstly. The resulting picture of our analysis indeed looked quite complicated, but the choices of the potentials and Lagrange multipliers were chosen in such a way so that concordance with observations is achieved. It is conceivable that the form of the $F(R)$ gravity was chosen abstractly and for illustrative purposes only. In principle, it is possible that the $F(R)$ gravity is chosen is such a way so that the resulting potentials can significantly be simplified. For example, we could choose the $F(R)$ gravity to be, 
\begin{equation}\label{frgravitychoiceforrev}
F(R)=R^{1+\epsilon}\, ,
\end{equation}
with the parameter $\epsilon$ being quite small. This kind of Einstein gravity modification was firstly considered in \cite{barrowearly}. For the $F(R)$ gravity chosen as in Eq. (\ref{frgravitychoiceforrev}), and for the Hubble rate of Eq. (\ref{hub2}), the resulting potential is,
\begin{equation}\label{revprd1}
V(N)\simeq 3 \left(-e^{\beta N} G_0+G_1\right)^2 \left(4+\left(-e^{\beta N} G_0+G_1\right)^2\right)-2 \left(e^{\beta N} G_0-G_1\right) \left(-3 G_1+e^{\beta N} G_0 (3+\beta )\right)\, ,
\end{equation}
where Eq. (\ref{revprd1}) results by expanding the original relation in terms of $\epsilon$ and keeping the leading order terms. So by comparing Eqs. (\ref{revprd1}) and (\ref{podnegsa}), it is obvious that the potential (\ref{revprd1}) is much less complicated in comparison to that of Eq. (\ref{podnegsa}). The same applies for the Lagrange multiplier $\lambda (N)$ which at leading order is,
\begin{equation}\label{revprd2}
\lambda (N)\simeq -3 \left(-e^{\beta N} G_0+G_1\right)^2-\frac{3}{2} \left(-e^{\beta N} G_0+G_1\right)^4+3 e^{\beta N} G_0 \left(e^{\beta N} G_0-G_1\right) \beta\, ,
\end{equation}
which has a much simpler form in comparison to that of Eq. (\ref{lajdbfe}). Hence, in principle it is possible to have a less complicated final form of the potential and Lagrange multiplier. Also this can be possible by appropriately choosing the Hubble rate, with only one constraint, to yield concordance with observational data. 

However, apart from the quantitative reasoning we just presented, the question still persists, how it is possible to single out the correct model. This question is deeper since it is related to the fundamental question-approach-discipline of physics, simplicity of a physical theory. Particular, it is an unwritten law that simplicity is one of the most compelling characteristics of a physical theory. If this discipline has to be respected, then the work we presented in this paper is just a mathematical exercise and no insights can be gained from our results. However, before all the physical phenomena related with cosmology are answered, it is not wise to undermine any theoretical approach. This because the physical results yet to be explained theoretically, cannot be explained with the physical theories at hand, so generalized theories are needed. In addition, a study can answer a question in a positive or a negative way. For example, the fact that the mimetic $F(R)$ gravity yields quite complicated potentials, can be regarded as a disadvantage of the theory, if simplicity should to be respected as a rule. Nevertheless, the theoretical approach we presented yields compatible results with the observed values of the spectral index of primordial curvature perturbations, and also with the values of the scalar-to-tensor ratio. So from this point of view, this theoretical approach is acceptable, however the comparison with observation we performed is not exhaustive, meaning that we did not take into account all the observational data, like for example implied by the theory non-Gaussianities, growth index of linear perturbations e.t.c. But from a theoretical point of view, the model we presented has quite appealing features, which can be listed as follows:
\begin{itemize}
    \item In the context of our model, the unification of early with late-time acceleration is achieved, and also a description for dark matter is provided too, without the need for the use of a perfect fluid dark matter sector.
    \item Our theory has an inherent conformal symmetry, which is one of the most fundamental symmetries in the Universe.
    \item The model is compatible with recent observational data.
\end{itemize}    
Thus, by searching for different modified gravities, we also try to search for best cosmological description fitted by observations. The reason for finding such complicated resulting potentials, may probably indicate that we eventually missed some fundamental symmetry of the Universe, so we should search among the models to find indications to it. On the other hand, complicated potentials indicate that not every cosmological description is permitted. In conclusion, our results indicate that the mimetic $F(R)$ theory must be further studied with two guidelines: simplicity and compatibility with observations. In addition, apart from the cosmological observational data, the mimetic $F(R)$ theory should be confronted with data coming from local astrophysical data. Also the implications of mimetic $F(R)$ gravity on compact objects, like neutron stars,should be scrutinized in future theoretical studies.

Coming back to the simplicity issue, it is definitely of great importance to combine simplicity with rigidity in theoretical descriptions of nature. However, many times the simplicity of an approach is a subject rather debatable, especially when one deals with the Universe, its origin and its evolution. But definitely, the resulting fundamental physical theory governing the cosmos should be simple.

In our way of thinking, bringing together all possible theoretical descriptions and then deciding which description is closer to reality, is also very important. The full theoretical understanding of the cosmic evolution also requires different theoretical approaches that can describe successfully the cosmic evolution. Then, a direct comparison of the various theoretical models can favor some models against some other theoretical models. Our approach showed that we can have viable cosmological evolution, compatible with Planck data, at the cost of having quite complicated potentials and Lagrange multipliers. Clearly, this could work as an indication that nature might be simpler than this, but in any case this should be scrutinized by examining all possible models. However, we presented some complicated models and perhaps a more plausible choice could result into much simpler potentials as we showed in Eqs. (\ref{revprd1}) and (\ref{revprd2}). Also, the complexity of the resulting theory might be an indication that nature does not actually favor the mimetic approach. But in order to reach a conclusion, the analysis we performed should be further scrutinized by focusing on astrophysical systems in order to see if the complex potentials structure persists. Such an indication will signal a problematic theory. For instance, we are finishing the study of Mass-Radius relation for neutron stars in frames of mimetic $F(R)$ gravity under discussion. Preliminary results indicate that viable models which we presented here, yield observationally preferable Mass-Radius relation for very compact neutron stars with large mass (about two Solar masses) in comparison to General Relativity.

We believe that indeed simplicity is nature's way, but even complexity is valuable since it may eventually indicate if an approach is plausible or not. However, this issue can be debatable, so we refer from going into further discussions on this.

\section{Conclusions}

In this paper we presented a new theoretical framework that enables us to
realize any arbitrary cosmology in a consistent and relatively easy way. The
theory is mimetic $F(R)$ gravity, and in addition to the Jordan frame $F(R)$
gravity, it is accompanied by a potential and a Lagrange multiplier. The
presence of the potential and Lagrange multiplier offers many different
possible ways for the realization of various cosmological evolutions. Of
particular interest is to try to produce cosmologies that are in concordance
with observations, and we started our analysis with this kind of models.
Given the $F(R)$ gravity, which can be arbitrarily chosen, we used the
mimetic $F(R)$ gravity formalism as a reconstruction method in order to find
which mimetic potential and Lagrange multiplier can produce such a
cosmology. The attribute of this reconstruction method is that the $F(R)$
gravity can be arbitrarily chosen, so one can use a viable $F(R)$ model,
which combines the following features: Firstly a successful and unified
description of early and late-time acceleration within the same model and
secondly, a consistent and elegant description of dark matter. Elegant
because there is no need for a particle description of dark matter, since it
is contained in the gravitational sector of the Jordan frame Lagrangian.

In addition to this study, a compelling task is to check if there exist de
Sitter points in the mimetic $F(R)$ gravity framework and also if these are
stable or unstable towards linear perturbations. In principle, this strongly
depends on the $F(R)$ gravity details, so we made an investigation for some
well known $F(R)$ models. As we demonstrated, there exist stable and
unstable de Sitter vacua in the models we studied. The unstable vacua are
very important from a physical point of view, since it is exactly these
vacua that lead to curvature perturbations which may realize a graceful exit
from inflation. The stable de Sitter vacua are the final attractors of the
trajectories of the dynamical system, and therefore in such a case,
inflation is eternal. However, as we also stressed in the text, in general the de Sitter solution instability, apart from graceful exit, may affect the primordial power spectrum, and consequently affect the predictions of the theory with regards to the CMB data. This issue is quite serious and should be appropriately addressed in a future work. We intend to report on this issue soon.

Furthermore, in the mimetic $F(R)$ gravity framework it is possible to
investigate which $F(R)$ gravities can realize certain bouncing cosmologies,
given the mimetic potential and Lagrange multiplier. The interesting feature
of this study is that the resulting $F(R)$ gravity is totally different from
the one that results in the pure $F(R)$ gravity case. In addition, since the
potential can be freely chosen, there is more than one $F(R)$ gravity that
can realize the same bouncing cosmology, so practically one has a class of
solutions for different potential and Lagrange multiplier.

We also performed an analysis of the Einstein frame counterpart theory
corresponding to the Jordan frame mimetic $F(R)$ gravity theory. As we
demonstrated, this theory is a two scalar field theory, with coupled kinetic
term for the mimetic auxiliary scalar field. Interestingly enough, it is
possible to produce models with appealing phenomenological properties, like
the one studied in \cite{linde}. We applied our findings for a Jordan frame
$R^2$ theory, for which we calculated the spectral index of primordial
perturbations and the running of the spectral index, when this is considered
in the Einstein frame. We used the two scalar field formalism of Refs.
\cite{kaizer}, and we also discussed the slow-roll limiting cases of this
formalism but also the non-slow-roll limit approximation for the same
theory.

Finally, we analyzed in detail the dynamical system that corresponds to the
mimetic $F(R)$ theory. By introducing new variables, we investigated which
are the fixed points of the dynamical system and studied their stability
against linear perturbations. Moreover, the physical significance of some of
the fixed points of the system was qualitatively discussed.

The mimetic $F(R)$ theory theoretical framework and the two reconstruction
methods it implies, offer many possibilities for realizing any cosmological
evolution, by using an arbitrary in principle $F(R)$ gravity. A particularly
interesting implication of the Jordan frame mimetic $F(R)$ gravity, is the
multi-scalar Einstein frame counterpart theory. In this paper we provided
the general theoretical framework and we also studied some particularly
interesting models. But there are many options for model building in
multi-scalar cosmology, so it is worth to further investigate the theory in
this direction and also perform a dynamical system analysis of the resulting
picture. Also, the slow-roll regime of the Einstein frame theory has to be
studied in a detail and in addition, even the theoretical study of the
non-slow-roll evolution (see for example \cite{barrowslowroll}) of both the
Jordan and Einstein frame theories, should be studied in some detail. We
need to stress that multi-scalar models for inflation are not favored by the
latest Planck observations \cite{planck}, however small deformations of
single scalar field models can be achieved by using for example two scalar
field models. For example, in the trajectory space of the solutions of the
model we examined in this paper, a small deformation of the single
Starobinsky model could be achieved by choosing the trajectory with
$\dot{\phi}\simeq 0$, $\phi\simeq 0$. In this way we could have all the
attributes of a single scalar field model, but we could also achieve
non-Gaussianities caused by the second scalar field. We hope to address
these issues in a future publication.

Finally, an interesting possibility for future applications related to the present paper, is to address the issues we discussed in this paper, in the context of another modified gravity theory, $F(T)$ gravity, see \cite{Cai:2015emx} for a recent compact study on the subject, and see also \cite{nashed}. In the $F(T)$ gravity approach, the scientific viewpoint is different from the approach we adopted in this paper, since the fundamental variable is not the metric anymore, but the vielbein, nevertheless the two formalisms share a lot of common features, with regards to the phenomenological applications. Particularly interesting could be to investigate the perturbations issue in mimetic $F(R)$ gravity, and for a non-singular cosmology and compare the results with the existing results on the subject from the current literature \cite{Cai:2011tc}.

\section*{Acknowledgments}

This work is supported in part by MINECO (Spain), project FIS2013-44881
(S.D.O) and partly by Min. of Education and Science of Russia (S.D.O and
V.K.O).

\section*{Appendix A: Detailed Form of the Parameters Appearing in Main
Text}

In this Appendix, we give the detailed form of the parameters appearing in
the previous sections. We start off with the parameters $\mathcal{A}$,
$\delta_1$ and $\delta_2$ that appear in Eq. (\ref{solutionbigfr3}), which
are equal to,
\begin{align}\label{newparametersdeltaparameters}
& \mathcal{A}=\frac{H_{dS} \left(-1-24 d H_{dS}^2+432 d H_{dS}^4\right)}{2
\left(1+48 d H_{dS}^2-1296 d H_{dS}^4\right)}\, , \\ \notag &
\delta_1=\frac{1}{2 \left(-12 d H_{dS}^2+432 d H_{dS}^4\right)}\Big{(}-1-36
d H_{dS}^2+48 d H_{dS}^3+864 d H_{dS}^4-1728 d H_{dS}^5\\ \notag & -\sqrt{-4
\left(-12 d H_{dS}^2+432 d H_{dS}^4\right) \left(2 H_{dS}+96 d H_{dS}^3-2592
d H_{dS}^5\right)+\left(1+36 d H_{dS}^2-48 d H_{dS}^3-864 d H_{dS}^4+1728 d
H_{dS}^5\right)^2}\Big{)}\, , \\ \notag &
\delta_2=\frac{1}{2 \left(-12 d H_{dS}^2+432 d H_{dS}^4\right)}\Big{(}-1-36
d H_{dS}^2+48 d H_{dS}^3+864 d H_{dS}^4-1728 d H_{dS}^5\\ \notag & +\sqrt{-4
\left(-12 d H_{dS}^2+432 d H_{dS}^4\right) \left(2 H_{dS}+96 d H_{dS}^3-2592
d H_{dS}^5\right)+\left(1+36 d H_{dS}^2-48 d H_{dS}^3-864 d H_{dS}^4+1728 d
H_{dS}^5\right)^2}\Big{)}\, .
\end{align}
In addition, the parameters $\zeta_1$ and $\zeta_2$ appearing in Eq.
(\ref{solutionbigfr312}), are equal to,
\begin{align}\label{newparameters}
& \zeta_1=\frac{-24 H_{dS}^2+2^{1+2 n} 3^n \left(H_{dS}^2\right)^n n \mu
+2^{2+2 n} 3^n H_{dS} \left(H_{dS}^2\right)^n n \mu -2^{2+2 n} 3^n H_{dS}
\left(H_{dS}^2\right)^n n^2 \mu +\sqrt{f(H_{dS})}}{2 \left(-2^{2 n} 3^n
\left(H_{dS}^2\right)^n n \mu +2^{2 n} 3^n \left(H_{dS}^2\right)^n n^2 \mu
\right)}
\, , \\ \notag &
\zeta_2=\frac{-24 H_{dS}^2+2^{1+2 n} 3^n \left(H_{dS}^2\right)^n n \mu
+2^{2+2 n} 3^n H_{dS} \left(H_{dS}^2\right)^n n \mu -2^{2+2 n} 3^n H_{dS}
\left(H_{dS}^2\right)^n n^2 \mu -\sqrt{f(H_{dS})}}{2 \left(-2^{2 n} 3^n
\left(H_{dS}^2\right)^n n \mu +2^{2 n} 3^n \left(H_{dS}^2\right)^n n^2 \mu
\right)}
\, .
\end{align}
where $f(H_{dS})$ stands for,
\begin{align}\label{sjdhd}
 & f(H_{dS})=-4 \left(48 H_{dS}^3-2^{2+2 n} 3^n H_{dS}
\left(H_{dS}^2\right)^n n \mu \right) \left(-2^{2 n} 3^n
\left(H_{dS}^2\right)^n n \mu +2^{2 n} 3^n \left(H_{dS}^2\right)^n n^2 \mu
\right) \\ \notag &
+\left(24 H_{dS}^2-2^{1+2 n} 3^n \left(H_{dS}^2\right)^n n \mu -2^{2+2 n}
3^n H_{dS} \left(H_{dS}^2\right)^n n \mu +2^{2+2 n} 3^n H_{dS}
\left(H_{dS}^2\right)^n n^2 \mu \right)^2
\end{align}
Finally, the parameters $\gamma_1$ and $\gamma_2$ that appear in Eq.
(\ref{finalfrsolutiongfg}) are equal to,
\begin{align}\label{d1kommadelta2}
& \gamma_1=\frac{\sqrt{2-c^2} \left(-2 \sqrt{6} \sqrt{2-c^2}-\sqrt{6} c^2
\sqrt{2-c^2}-\sqrt{-528+600 c^2-156 c^4-6 c^6}\right)}{2 \sqrt{6} c^2
\left(-2+c^2\right)}\, , \\ \notag & \gamma_2=\frac{\sqrt{2-c^2} \left(-2
\sqrt{6} \sqrt{2-c^2}-\sqrt{6} c^2 \sqrt{2-c^2}+\sqrt{-528+600 c^2-156 c^4-6
c^6}\right)}{2 \sqrt{6} c^2 \left(-2+c^2\right)}\, .
\end{align}

\section*{Appendix B: The Detailed Form of the Parameter $M_{\sigma \sigma}$
for the Einstein Frame $R^2$ Potential}

In this Appendix, we provide the detailed form of the parameter $M_{\sigma
\sigma}$ of Eq. (\ref{msssdetailed}), in the case the potential $U(\varphi)$
is chosen as in Eq. (\ref{starobpot}). By substituting the potential of Eq.
(\ref{starobpot}) in Eq. (\ref{msssdetailed}), we obtain,
\begin{align}\label{mss}
& M_{\sigma \sigma}=-3 A C_0 e^{\sqrt{\frac{3}{2}} \kappa  \varphi } \kappa
^2+6 A C_0 e^{\sqrt{6} \kappa  \varphi } \kappa ^2-6 A e^{-\sqrt{6} \kappa
\varphi } \kappa ^2 V(\phi )-3 A C_0 \kappa ^2 \lambda (\phi ) \\ \notag &
-6 B C_0 \kappa ^2 \lambda (\phi )+3 A C_0 e^{-\sqrt{\frac{3}{2}} \kappa
\varphi } \kappa ^2 \lambda (\phi )+6 B C_0 e^{-\sqrt{\frac{3}{2}} \kappa
\varphi } \kappa ^2 \lambda (\phi )+6 A e^{-\sqrt{6} \kappa  \varphi }
\kappa ^2 \lambda (\phi )-3 C_0 \Delta  \kappa ^2 \lambda (\phi )+3 C_0
e^{-\sqrt{\frac{3}{2}} \kappa  \varphi } \Delta  \kappa ^2 \lambda (\phi )\\
\notag &-3 A e^{-2 \sqrt{6} \kappa  \varphi } \kappa ^2 V(\phi ) \lambda
(\phi )-6 B e^{-2 \sqrt{6} \kappa  \varphi } \kappa ^2 V(\phi ) \lambda
(\phi )-3 e^{-2 \sqrt{6} \kappa  \varphi } \Delta  \kappa ^2 V(\phi )
\lambda (\phi )+3 A e^{-2 \sqrt{6} \kappa  \varphi } \kappa ^2 \lambda (\phi
)^2+6 B e^{-2 \sqrt{6} \kappa \varphi } \kappa ^2 \lambda (\phi )^2\\ \notag
&+3 e^{-2 \sqrt{6} \kappa \varphi } \Delta  \kappa ^2 \lambda (\phi
)^2+\sqrt{\frac{3}{2}} A e^{-\sqrt{6} \kappa  \varphi } \kappa  V'(\phi )+3
\sqrt{6} B e^{-\sqrt{6} \kappa  \varphi } \kappa  V'(\phi
)-\sqrt{\frac{3}{2}} A e^{-\sqrt{6} \kappa  \varphi } \kappa \lambda '(\phi
)-3 \sqrt{6} B e^{-\sqrt{6} \kappa  \varphi } \kappa  \lambda '(\phi )\\
\notag &+\frac{e^{-\sqrt{6} \kappa  \varphi } \Delta  V'(\phi ) \lambda
'(\phi )}{2 \lambda (\phi )}-\frac{e^{-\sqrt{6} \kappa  \varphi } \Delta
\lambda '(\phi )^2}{2 \lambda (\phi )}-e^{-\sqrt{6} \kappa  \varphi } \Delta
V''(\phi )+e^{-\sqrt{6} \kappa  \varphi } \Delta  \lambda ''(\phi )\, .
\end{align}

\section*{Appendix C: Useful Relations for the Derivation of the Dynamical
System}

In this appendix we quote some useful relations necessary for the derivation
of the dynamical system of Eq. (\ref{dynamicalsystemmain}). These are the
following,
\begin{equation}\label{r1}
-\frac{\ddot{F}}{Fh^2}=-2x_1-3x_4-x_3x_6+2x_6-3x_6-3x_2-3x_4\, ,
\end{equation}
\begin{equation}\label{r2}
\dot{F}=F'(R)\left( 6 \ddot{H}+4H\dot{H}\right)\, ,
\end{equation}
\begin{equation}\label{r3}
\dot{V}=-12H\lambda-6\dot{H}\dot{F'}-6H\ddot{F'}+6\left
(\ddot{H}+2H\dot{H}\right)F'+6\left(\dot{H}+H^2\right)-\dot{F}\, ,
\end{equation}
\begin{equation}\label{r4}
\dot{\lambda}=-\frac{\dot{F}}{2}+3\left(\ddot{H}+2H\dot{H}\right)F'+3\left(\
dot{H}+H^2\right)\dot{F'}-3\dot{H}\dot{F'}-3H\ddot{F'}\, ,
\end{equation}
where in all the above equations, the prime denotes differentiation with
respect to $R$ and we have assumed the identification $\phi=t$.

\section*{Appendix D: The Dynamical System in the Presence of Perfect Matter
Fluid}

Here we quote the exact form of the fixed points corresponding to the
dynamical system of Eq. (\ref{dynamicalsystemmainwithmatter}), which are,
\begin{align}\label{fixedpointswm}
& P_1: x_1\to 0,\,\,\,\,\,x_2\to
-2+\frac{\sqrt{m}}{\sqrt{2}},\,\,\,\,\,x_3\to \frac{1}{2} \left(4-\sqrt{2}
\sqrt{m}\right),\,\,\,\,\,x_4\to \frac{-414 \sqrt{2} \sqrt{m}+60 m+79
\sqrt{2} m^{3/2}-22 m^2}{162-45 m+2 m^2},\\ \notag & x_5\to \frac{1080-90
\sqrt{2} \sqrt{m}-192 m+33 \sqrt{2} m^{3/2}-2 m^2}{162-45 m+2
m^2},\,\,\,\,\,x_6\to \frac{3 \left(108+405 \sqrt{2} \sqrt{m}+45 m-64
\sqrt{2} m^{3/2}+2 m^2\right)}{162-45 m+2 m^2},\,\,\,x_7\rightarrow 0\, ,
\end{align}
\begin{align}\label{fixedpoints1wm}
& P_2: x_1\to 0,\,\,\,\,\,x_2\to \frac{1}{2} \left(-4-\sqrt{2}
\sqrt{m}\right),\,\,\,\,\,x_3\to \frac{1}{2} \left(4+\sqrt{2}
\sqrt{m}\right),\,\,\,\,\,x_4\to \frac{414 \sqrt{2} \sqrt{m}+60 m-79
\sqrt{2} m^{3/2}-22 m^2}{162-45 m+2 m^2},\\ \notag & x_5\to \frac{1080+90
\sqrt{2} \sqrt{m}-192 m-33 \sqrt{2} m^{3/2}-2 m^2}{162-45 m+2
m^2},\,\,\,\,\,x_6\to \frac{3 \left(108-405 \sqrt{2} \sqrt{m}+45 m+64
\sqrt{2} \, . m^{3/2}+2 m^2\right)}{162-45 m+2 m^2},\,\,\,x_7\rightarrow 0\,
,
\end{align}
\begin{align}\label{strangeones1}
& P_3: x_1\to 3-\sqrt{2} \sqrt{m}+3 w,\,\,\,\,\,x_2\to \frac{-2 \sqrt{2}
\sqrt{m}+m}{3 (1+w)},\,\,\,\,\,x_3\to \frac{1}{2} \left(4-\sqrt{2}
\sqrt{m}\right),\\ \notag &
x_4\to \frac{1}{3 \left(-3+6 w^2+4 w^3+w^4\right)}\times \\ \notag &
\Big{(}15-\frac{23 \sqrt{m}}{\sqrt{2}}+4 m+70 w-21 \sqrt{2} \sqrt{m} w+2 m
w+89 w^2-18 \sqrt{2} \sqrt{m} w^2-3 m w^2\\ \notag &
19 w^3+3 \sqrt{2} \sqrt{m} w^3-3 m w^3-24 w^4+\frac{15 \sqrt{m}
w^4}{\sqrt{2}}-9 w^5\Big{)}\, ,
\\ \notag & x_5\to \frac{1}{6 \left(-3+3 w+3 w^2+w^3\right)}\times \\ \notag
&
\Big{(}6-10 \sqrt{2} \sqrt{m}+10 m+490 w-169 \sqrt{2} \sqrt{m} w+12 m w+414
w^2-111 \sqrt{2} \sqrt{m} w^2\\ \notag &
2 m w^2+120 w^3-19 \sqrt{2} \sqrt{m} w^3+6 m w^3+48 w^4-15 \sqrt{2} \sqrt{m}
w^4+18 w^5
\Big{)}
\\ \notag & x_6\to 0,\,\,\,x_7\rightarrow \frac{1}{6 \left(-3+6 w^2+4
w^3+w^4\right)}\times \\ \notag &
\Big{(}276-93 \sqrt{2} \sqrt{m}+18 m+660 w-234 \sqrt{2} \sqrt{m} w+24 m
w+522 w^2-152 \sqrt{2} \sqrt{m} w^2\\ \notag &
8 m w^2+186 w^3-34 \sqrt{2} \sqrt{m} w^3+6 m w^3+66 w^4-15 \sqrt{2} \sqrt{m}
w^4+18 w^5
\Big{)}
\end{align}
\begin{align}\label{strangeones2}
& P_4:: x_1\to 3+\sqrt{2} \sqrt{m}+3 w,\,\,\,\,\,x_2\to \frac{2 \sqrt{2}
\sqrt{m}+m}{3 (1+w)},\,\,\,\,\,x_3\to \frac{1}{2} \left(4+\sqrt{2}
\sqrt{m}\right),\\ \notag &
x_4\to \frac{1}{6 \left(-3+6 w^2+4 w^3+w^4\right)}\times \\ \notag &
\Big{(}30+23 \sqrt{2} \sqrt{m}+8 m+140 w+42 \sqrt{2} \sqrt{m} w+4 m w+178
w^2+36 \sqrt{2} \sqrt{m} w^2\\ \notag &
6 m w^2+38 w^3-6 \sqrt{2} \sqrt{m} w^3-6 m w^3-48 w^4-15 \sqrt{2} \sqrt{m}
w^4-18 w^5\Big{)}\, ,
\\ \notag & x_5\to \frac{1}{9-9 w-9 w^2-3 w^3}\times \\ \notag &
\Big{(}-3-5 \sqrt{2} \sqrt{m}-5 m-245 w-\frac{169 \sqrt{m} w}{\sqrt{2}}-6 m
w-207 w^2-\frac{111 \sqrt{m} w^2}{\sqrt{2}}-m w^2\\ \notag &
60 w^3-\frac{19 \sqrt{m} w^3}{\sqrt{2}}-3 m w^3-24 w^4-\frac{15 \sqrt{m}
w^4}{\sqrt{2}}-9 w^5
\Big{)}
\\ \notag & x_6\to 0,\,\,\,x_7\rightarrow \frac{1}{9-18 w^2-12 w^3-3
w^4}\times \\ \notag &
\Big{(}-138-\frac{93 \sqrt{m}}{\sqrt{2}}-9 m-330 w-117 \sqrt{2} \sqrt{m}
w-12 m w-261 w^2-76 \sqrt{2} \sqrt{m} w^2\\ \notag &
4 m w^2-93 w^3-17 \sqrt{2} \sqrt{m} w^3-3 m w^3-33 w^4-\frac{15 \sqrt{m}
w^4}{\sqrt{2}}-9 w^5
\Big{)}\, .
\end{align}
As we can see, the points $P_1$ and $P_2$ coincide with the fixed points of
the dynamical system (\ref{dynamicalsystemmain}), while $P_3$ and $P_4$
contain the point $x_6\rightarrow 0$, which can be considered unphysical,
unless the $F(R)$ gives such a condition for a suitable chosen $F(R)$
function, for all $N$.

\end{document}